\documentclass[usenatbib,useAMS]{mn2e}
\usepackage{url}
\usepackage{threeparttable}
\usepackage{multirow}
\usepackage{graphicx}
\usepackage{pdflscape}
\usepackage[caption=false]{subfig}
\usepackage{afterpage}

\usepackage{amssymb}
\usepackage{amsmath}

\mathchardef\mhyphen="2D

\title{Galaxy And Mass Assembly: Search for a population of high entropy galaxy groups}
\author[Pearson, et al.]{R. ~J. ~Pearson,$^1$\thanks{Email: richard@star.sr.bham.ac.uk}
		T. ~J. ~Ponman,$^1$\thanks{Email: tjp@star.sr.bham.ac.uk} 
		P. ~Norberg,$^2$ A. ~S. ~G. ~Robotham,$^3$
		\newauthor A. ~Babul,$^4$ R. ~G. ~Bower,$^2$ I. ~G. ~McCarthy,$^5$ S. ~Brough,$^6$ S. ~P. ~Driver,$^{3,7}$ 
		\newauthor K. ~Pimbblet$^{8,9}$  \\
		$^1$ School of Physics and Astronomy, University of Birmingham, Edgbaston, Birmingham B15 2TT, UK \\
		$^2$ ICC and CEA, Department of Physics, Durham University, South Road, Durham DH1 3LE, UK \\
		$^3$ International Centre for Radio Astronomy Research, University of Western Australia, 35 Stirling Highway, Crawley, WA 6009, Australia \\
		$^4$ Department of Physics \& Astronomy, University of Victoria, BC V8X 4M6, Canada \\
		$^5$ Astrophysics Research Institute, Liverpool John Moores University, 146 Brownlow Hill, Liverpool L3 5RF, UK \\
		$^6$ Australian Astronomical Observatory, PO Box 915, North Ryde, NSW 1670, Australia \\
		$^7$ School of Physics \& Astronomy, University of St Andrews, North Haugh, St Andrews KY16 9SS, UK \\
		$^8$ E.A.Milne Centre for Astrophysics, University of Hull, Cottingham Road, Kingston-upon-Hull, HU6 7RX, UK \\
		$^9$ School of Physics and Astronomy, Monash University, Clayton, Victoria 3800, Australia
		}

\begin{document}

\maketitle
\begin{abstract}

	Observations with the {\it Chandra} X-ray Observatory are used to examine
        the hot gas properties within a sample of 10 galaxy groups selected
        from the Galaxy And Mass Assembly survey's optical friends-of-friends group catalogue.
        Our groups have been screened to eliminate spurious and unrelaxed systems, and the 
        effectiveness of this procedure is demonstrated by the detection of intergalactic
        hot gas in $80$ per cent of our sample. However we
        find that 9 of the 10 are X-ray underluminous by a mean factor $\sim 4$ compared to typical X-ray
        selected samples. Consistent with this, the majority of our groups have gas fractions	which are lower and gas entropies somewhat higher than those seen in typical 
        X-ray selected samples. Two groups, which have high $2\sigma$ lower 
        limits on their gas entropy, are candidates for the 
        population of high entropy groups predicted by some AGN feedback models.

\end{abstract}

\section{Introduction} \label{SINTRO}

Studies of the hot gas in groups and clusters of galaxies have demonstrated that
this gas exhibits entropies in excess of self-similar expectations
\citep[e.g.][]{PONMAN99,LLOYDDAVIES00}.  Determining how, why and when
this entropy was raised is essential to better understand the
formation and evolution of both galaxies and galaxy clusters. For
example, the processes responsible for raising the gas entropy may be
the same as those which maintain the hot gas content of the Universe
by preventing runaway cooling of gas into stars \citep[the cooling
catastrophe, ][]{BALOGH01}.

Early models suggested that entropy had been injected into the
intracluster medium (ICM) prior to the full collapse of the cluster
\citep{KAISER91,EVRARD91}. Proposed sources of entropy included
supernova feedback and active-galactic nuclei \citep{LLOYDDAVIES00,
WU00, TOZZI01, BABUL02}.  Due to the smaller potential wells of groups compared
to massive clusters, the entropy injected into gas at group scales can
be significant compared to the entropy accrued during halo
assembly. Alternatives, such as the action of cooling flows, which 
preferentially cool low entropy gas into stars, raising the
\emph{mean} ICM entropy \citep[see, e.g.][]{BRYAN00, VOIT01}, have
also been proposed.  It is now widely recognised that feedback from Active Galactic Nuclei (AGN)
reduces the cooling rate within cluster cores, preventing massive cooling flows.
Such feedback after cluster collapse provides another potential mechanism for raising the 
entropy of the ICM \citep{VOIT05REV}.

Both analytical models \citep{MCCARTHY04, MCCARTHY08} and hydrodynamical simulations \citep{LEWIS00, DAVE08, SCHAYE09, LIANG16} have been used to
explore the evolution of gas within group and cluster halos,
examining the effect of various proposed feedback models. Comparison between the simulations
from the Overwhelmingly Large Simulations Project \citep[OWLS, ][]{SCHAYE09} 
and a variety of observational properties of groups \citep{MCCARTHY10,MCCARTHY11}
suggest the observed raised entropies result from heating of gas by AGN feedback
acting within precursor halos. These models predict a distribution 
of central entropies broader than observed in the well-studied X-ray selected
samples of groups, and in particular the existence of a population of
systems with very high central entropy. Such high entropy groups would have low
central gas densities, and hence low surface brightness X-ray emission. As a result,
systems of this sort would be heavily selected against in X-ray bright samples.
The primary aim of the present study is to search for a population of galaxy groups
($M \lesssim 10^{14} \mathrm{M_{\sun}}$) containing such high entropy gas.

The starting point for such a search has to be a group sample which has been
constructed without any reference to X-ray properties. We therefore start from an
optically selected sample of groups. Due to the depth and completeness of the
Galaxy and Mass Assembly \citep[GAMA, ][]{DRIVER09a, LISKE15} project's spectroscopic survey we draw 
our group candidates from their Friends-of-Friends group catalogue \citep{ROBOTHAM11}. The difficulty with
such optically selected groups is to avoid contamination of the sample by
chance galaxy alignments or by halos that have yet to fully collapse. Such systems
would have little or no diffuse X-ray emission, and so could be mistaken
for high entropy groups. We therefore apply a number of tests designed to eliminate
such spurious or unrelaxed systems from our sample.

The outline of this paper is as follows: In \S\ref{SDATA} we discuss how we select our sample
of relaxed optical groups by testing for substructure in the distribution of member 
galaxies. \S\ref{SXANALYSIS} describes the X-ray analysis performed on the {\it Chandra} data and 
\S\ref{SRESULTS} presents the results for the X-ray luminosity ($L_X$), and the mass and
central entropy of the hot gas within our groups. We
conclude, in \S\ref{SDISCUSSION} and \S\ref{SCONCLUSION}, by discussing the virialisation
of these groups and the limitations in our analysis which could affect our entropy constraints.
Throughout this paper we adopt a simple $\Lambda$CDM cosmology with $\Omega_m = 0.3$, $\Omega_\Lambda = 0.7$
and $H_0=70 \mathrm{~km ~s^{-1} ~Mpc^{-1}}$, with $h=H_0/100 \mathrm{~km ~s^{-1} ~Mpc^{-1}}=0.7$.

\section{The Optical Group Sample} \label{SDATA}

To investigate whether a population of high entropy groups exists we
need to start from a sample of relaxed, optically selected groups. 
As discussed above, this
removes ambiguity in the nature of any observed low $L_X$ groups. The
GAMA survey provides an excellent platform from which to
begin a study of this nature.

The Galaxy And Mass Assembly project is a broad multi-wavelength
survey covering $\sim 290 \deg^2$ of the sky. The main optical
component of the project is a medium-deep redshift survey conducted with
the AAOmega multi-object spectrograph at the Anglo-Australian
Observatory.  This provides spectra and redshifts \citep{HOPKINS13} for more than
300,000 galaxies within five sky regions.  A grouping analysis has
been performed on the galaxies from the first three regions surveyed 
by \citet{ROBOTHAM11} using a
Friends-of-Friends (FoF) algorithm optimised on a sample of 9
realistic mock light cones \citep{MERSON13}.  These light cones were
generated using the dark matter Millennium Simulation
\citep{SPRINGEL05} populated with galaxies using the \citet{BOWER06}
semi-analytic models of galaxy formation, with modifications to
reproduce the observed $r$-band, redshift-dependent, GAMA luminosity
function of \citet{LOVEDAY12}.

At the time the present study began, most of the GAMA survey was
complete to an $r$-band magnitude of only $m_r=19.4 ~\mathrm{mag}$, and the G3Cv04 group
catalogue was constructed from these data by \citet{ROBOTHAM11}. Since
then, additional galaxy spectra have been taken, extending the
completeness to $m_r = 19.8 ~\mathrm{mag}$ and an updated (G3Cv06) group catalogue has
been compiled. Our selection of a relaxed group sample, as described in Section \ref{SDATA}, makes use of the G3Cv04
group catalogue, and {\it Chandra} data were acquired for the resulting sample. All
subsequent analysis presented here beyond the inital
selection makes use of the deeper data available from the G3Cv06 group catalogue.

The G3Cv04 catalogue contained 12,200 FoF groups and clusters with two or more
members brighter than $m_r = 19.4 ~\mathrm{mag}$.  We first cut the catalogue to restrict it to groups with $N
\geq 12$ galaxies for two reasons: firstly this essentially eliminates
the possibility of spurious groups generated from chance alignment of
physically unrelated galaxies, and secondly it ensures that we have
reasonable galaxy statistics for the substructure analysis we
subsequently applied to filter out unrelaxed systems. This initial cut
left a catalogue of 205 galaxy groups and clusters.  We then restricted the sample
to redshifts $z \leq 0.12$, since more distant systems passing our
membership criterion are fairly rich clusters, whilst our focus in
this study is on galaxy groups ($M_{500} \lesssim 10^{14} \mathrm{M_{\sun}}$).
This redshift cut reduced the sample to 64 galaxy groups which were subjected 
to the substructure screening described below. We found these limits on richness
and redshift provided the best compromise between the conflicting requirements
of increasing the number of candidate groups to select from whilst ensuring statistically
useful numbers of group members and reducing the likelihood of including clusters.

\subsection{Selection} \label{SDATASEL}

In order to be confident that low X-ray luminosity in any of
our groups is indicative of high entropy gas, it is important
to restrict our sample to systems which are collapsed and dynamically
relaxed. To do this, we applied a number of tests to assess
their degree of substructure. A highly
substructured group is likely to be collapsing for the first time,
or to have suffered a recent major merger. 

Another advantage of relaxed systems is that their masses can be
more reliably estimated, and should be closely related to their
X-ray properties. Since we 
use optical mass estimates to predict the X-ray properties of our
groups, reliable mass estimates are important
for correctly judging the exposure times required to acquire X-ray
data of sufficient depth for our study. It is especially important
to avoid selecting groups whose masses are overestimated, which would
result in predicted exposure times that would be too short.

The substructure statistics employed were calibrated on mock G3Cv04 data
constructed for the GAMA project.  Using the mock halos and with known masses and
masses predicted from their optical luminosities (see Section \ref{SGRPSAMPLE}),
we tune the selection to discriminate
against groups whose mass estimates are more than a factor of 2 larger
than the true halo mass.

The selection and calibration of the algorithms we used proceeded
as follows. First, we
identified a set of substructure statistics that had useful
discriminating power to select mock groups with `good' predicted masses.
Secondly, we established a set of thresholds
for the adopted statistics that, in combination, maximized the
number of mock groups recovered with satisfactory mass estimates. Due
to the importance of rejecting groups with overestimated masses, we
insisted that our screening process should essentially eliminate
any systems whose masses were overestimated by more than a factor of 2.
Finally, having tuned our selection criteria on mock
groups, we applied this filter to the real G3Cv04 sample.

\subsection{Substructure Statistics}

We initially examined nine different substructure statistics that 
probed the spatial and velocity distributions of the galaxies within a
group. Six of these were drawn from the compilation of
\citet{PINKNEY96}. Two more (an axial ratio and a group symmetry test)
were based upon the output of the
\citet{ROBOTHAM11} group analysis, with a further statistic used
specifically to examine the distribution of galaxy velocities. 
It was found that a combination of three of these tests was able
to effectively discriminate between systems for which the GAMA
mass estimates described in Section~\ref{SGRPSAMPLE} were reliable, and those for which
they were not. Adding further substructure tests to this set produced
negligible further gains. Our
chosen set of substructure tests consisted of two spatial symmetry
statistics: the $\beta_{\mathrm{sym}}$-test and the angular separation test
\citep{PINKNEY96}, together with the Anderson-Darling test for normality
\citep{THODE} in the velocity histogram, implemented through the
\textsc{nortest r} package \citep[][and references therein]{NORTEST} for the \textsc{r} statistical package \citep{R}. Two of these
tests, the $\beta_{\mathrm{sym}}$-test and the angular separation test, require an optical group centre. 
A centroid was calculated in each case by taking an unweighted centroid of group
galaxies within a radius which was iteratively shrunk until it encompassed $50$ per cent of a given group's
galaxies.

\subsubsection{The $\beta_{\mathrm{sym}}$ test}

This substructure test, discussed in \citet{WEST88} and
\citet{PINKNEY96}, looks for
deviations from mirror symmetry caused by substructure within a group
halo. The test proceeds by taking each galaxy and estimating the mean
separation, $d_i$, of the $\sqrt{N}$ nearest galaxies to it, 
where $N$ is the total number of galaxies within the cluster. This
distance is then compared to the same quantity calculated about a point
diametrically opposite to the galaxy, $d_o$, through a central point.

For the $i$th galaxy, $\beta_{\mathrm{sym}}$ is then defined as
\begin{equation}
 \beta_{\mathrm{sym},i} = \log \left(\frac{d_o}{d_i}\right) \quad .
\end{equation}

The mean value of $\beta_{\mathrm{sym}}$ for all galaxies is the unnormalised
$\beta_{\mathrm{sym}}$-statistic. For an unsubstructured, symmetric system $\beta_{\mathrm{sym}}
\approx 0$.

\subsubsection{The Angular Separation Test}

The angular separation test
\citep[hereafter AST, ][]{WEST88,PINKNEY96}, examines the projected angular
distribution of galaxies within the cluster, 
testing for an excess of small
angular separations which could indicate substructure. The AST first
determines the mean harmonic separation of members,

\begin{equation}
 \theta_{hm} = \left[\frac{2}{N(N-1)} \sum_{i=1;i>j}^N\theta_{ij}^{-1} \right]^{-1} \quad ,
\end{equation}

\noindent where $N$ is again the number of galaxies in the group or
cluster and $\theta_{ij}$ is the angular separation of two galaxies
relative to the centre of the group. Substructures such as infalling groups
should therefore reduce the value of $\theta_{hm}$ relative to a similar halo without any
substructure.

The values for both the $\beta_{\mathrm{sym}}$ test and for AST, are finally
normalised by their value from the ``null hypothesis'' -- a value of the
statistic when no substructure is present. This accounts for any
contributions to the measured statistic from statistical noise in the
population. To generate the null hypothesis, we perform the tests on
1000 realisations of the cluster data where the azimuthal positions of
the galaxies have been randomised. We then take the mean of these
ensembles to represent the substructure-less null hypothesis. The
final test statistics, $\zeta_{\beta}$ and $\zeta_{AST}$ are
\begin{equation}
 \zeta_{\beta} = \beta_{\mathrm{sym}} / \beta_{\mathrm{sym}, null} \quad \mathrm{and} \quad \zeta_{AST} = \theta_{null} / \theta_{hm} \quad .
\end{equation}

With this normalisation, substructure-less groups have statistic values
of $\approx 1$ whilst substructured systems will have statistics
$>1$. The significance of the statistic can be easily quantified
since it is the fraction of
the 1000 randomisations that have \emph{more} substructure than the
measured statistic.

\subsubsection{Anderson-Darling test}

Within a virialised structure, we expect the velocity of galaxies to
be distributed as a Gaussian along the line-of-sight. Recent cluster
merger activity or incomplete virialisation, as well as inclusion
of physically unconnected galaxies in a group,
would be expected to cause deviations from this
distribution, e.g. through introducing bimodality or asymmetries in
the velocity histogram. The Anderson-Darling test \citep[hereafter the
AD test, ][]{THODE} examines whether a sample is consistent with having
been drawn from a normal distribution, and can therefore be useful for
testing for such deviations. As a reasonable proxy for galaxy
velocities, we apply the test to the redshifts of member galaxies.

Here we describe the AD test as laid out by \citet{THODE} and as
implemented by the \textsc{nortest r} package \citep{NORTEST}. The AD
test proceeds by first taking the data, in this case, galaxy
redshifts, $z$, and sorting and scaling them relative to the mean, $\bar{z}$, and
standard deviation, $\sigma_z$, of the sample, 
\begin{equation}
 z^\prime = \frac{z-\bar{z}}{\sigma_z} \quad .
\end{equation}

The statistic, $A^2$, is then determined using the normal cumulative distribution
function $\Phi$ as,
\begin{equation}
 A^2 = -N - \frac{1}{N} \sum_i^N [2i-1] [\ln(\Phi(z_i^\prime)) + \ln(1-\Phi(z_{N-i+1}^\prime))]  ,
\end{equation}

\noindent where $N$ is the sample size. The significance of $A^2$, $p(A^2)$, 
can then be found as per Table 4.9 of \citet{DAGOSTINO} where a modified 
statistic, $A^{2\star}$, is adopted such that
\begin{equation}
 A^{2\star} = \left(1 + \frac{0.75}{N} + \frac{2.25}{N^2}\right)A^2 \quad .
\end{equation}

We use $p(A^2)$ as our substructure indicator. To be consistent with
the sense of
$\zeta_{\beta}$ and $\zeta_{AST}$, we define $\zeta_{AD} = 1 / p(A^2)$
such that a value $\approx 1$ indicates low levels of substructure, and
$\zeta_{AD} \gg 1$ shows substantial non-Gaussian structure in the
velocity histogram.

\subsection{Calibration} \label{SSUBCAL}

Using the mock galaxy groups we
calibrated a set of threshold values for our three
substructure statistics which selected out groups meeting the
criteria discussed in Section \ref{SDATASEL} for reliability of mass
estimation. To
ensure joint optimisation of the three substructure statistics, we
explored a gridded ($\zeta_{\beta}, \zeta_{AST}, \zeta_{AD}$) parameter
space, characterising each combination by the accuracy of the
predicted masses (see \ref{SGRPSAMPLE}) for the groups that passed the filter. Using one of
the GAMA mock volumes, we optimised the threshold values
to discard all groups with
predicted masses greater than twice the true halo mass whilst
maximising the number of groups with mass estimates within a factor of
two of the true mass. As each mock FoF group may link galaxies from multiple dark matter
halos, we take the true mass FoF group to be the mass of the halo that
contributes the most galaxies to the group.

The resulting calibration accepted groups with substructure measures
$\zeta_{\beta} < 1.9$, $\zeta_{AST} < 1.68$ and $\zeta_{AD} <
1.82$. Of the 62 groups with
$N_{FoF} \geq 12$ and $z \leq 0.12$ within the mock volume used
for the calibration these thresholds were able to
exclude all groups whose masses were overestimated by a factor of two
whilst allowing 17 per cent of the 42 groups with masses within a factor of
two (i.e.  `accurate' mass estimates) to pass. Using the same set of
threshold on the other eight other mock volumes, which contained 
in total 763 groups, we found that the filter allowed 16 per cent of
groups with `accurate' masses to pass whilst allowing only 6 per cent
contamination by groups whose masses are overpredicted by a factor
$\geq$2. Note that our filter will allow through some
groups whose masses are significantly {\it underestimated}.

\subsection{Group Sample} \label{SGRPSAMPLE}

Applying the calibrated substructure filters to the observed G3Cv04
group catalogue resulted in a sample of 18 groups with $N_{FoF} \geq
12$ and $z \leq 0.12$. Our aim was then to obtain, for as many of
these as possible, {\it Chandra} X-ray observations of a depth
sufficient to detect the intragroup gas, even if it has entropy higher
than that normally expected within a collapsed group. In order to
calculate the required X-ray exposure times we needed estimates of the
luminosity and temperature of the hot intragroup gas for each
target. For typical groups, both properties are found to correlate strongly with group
mass \citep[e.g.][]{SUN09}, so we use estimated group halo mass
as the basis for our prediction. Group mass has been found to
correlate well with optical luminosity
\citep[e.g.][]{POPESSO07}, so we adopted the group $r$-band luminosity 
as a predictor for group mass, using this in turn to predict the X-ray
properties. We use the relation
\begin{equation}
  \frac{M_{GAMA, DHalo}}{h^{-1} \mathrm{M_{\sun}}} = 10^{1.37} \left(\frac{L_r}{h^{-2} \mathrm{L_{\sun}}}\right)^{1.09} \quad , \label{EMGAMA2}
\end{equation}
where $L_r$ is the total $r$-band luminosity of the FoF group. This is estimated from
the observed $r$-band luminosity of the FoF-linked galaxies ($L_{obs}$) using
$L_r = B L_{obs} f(z)$, where $f(z)$ is an extrapolation
factor to account for GAMA's flux limit and $B=1.04$ is a correction factor
dependent on both group richness and redshift calibrated on the GAMA
mock catalogues \citep{ROBOTHAM11}. This quantity is available directly from the group
catalogue. We compare this relation to recent relations obtained by \citet{HAN15} and \citet{VIOLA15} who both determined mass-observable relations for GAMA groups using weak lensing. They find $M_{halo} \propto L_{r}^{1.08\pm0.22}$ and $M_{200} \propto L_{r}^{1.16\pm0.13}$ respectively, comparable to that used here.

Equation \eqref{EMGAMA2} was
calibrated using mock groups with dark halo masses $M_{DHalo}$
\citep{JIANG14, TANKARDEVANS15}. These luminosity-derived masses $M_{GAMA, DHalo}$ were
then converted into overdensity masses, $M_{500}$ (mass within the
region where mean density is 500 times the critical density of the 
Universe), using $M_{500} = 10^{0.34}M_{GAMA, DHalo}^{0.96}$, estimated from simulated
dark matter distributions \citep{JIANG14, TANKARDEVANS15}. 

The resulting values of $M_{500}$ were then used to predict $r_{500}$ radii, assuming a mean 
density $500$ times the critical density of the Universe at the group's redshift, and the
X-ray temperatures for each potential target group
using the mass-temperature relation of \citet{SUN09}. These temperatures
were then used to estimate X-ray luminosities
using the $L_X-T$ relation derived by \citet{SLACK14} for archival data. As we are interested in
high entropy groups, we would expect these groups to be underluminous
relative to the typical group population. Based on the simulations of
\citet{MCCARTHY10,MCCARTHY11} we anticipated that the highest
entropy groups might have X-ray luminosities suppressed by as much as
a factor of 10. We therefore calculated fluxes and {\it Chandra}
count rates assuming X-ray luminosities an order of magnitude below
the mean $L_X-T$ relation. From our sample of 18 potential target groups
we then selected the 10 groups with the shortest
exposure times required to reach a $3\sigma$ detection under these
constraints. One of these had existing archival {\it Chandra}
data sufficient for our purposes, and we were awarded observations of
the remaining nine.

We present our selected sample in Table \ref{TGROUPS}, where observed and
predicted group properties have been re-estimated using data from the
G3Cv06 group catalogue. Group 200130 is the well-known, relaxed X-ray group MKW4
\citep{FUKAZAWA04}. This was the group with existing archival
data, and was also the only group in our sample
that intersected the edge of a GAMA field. Approximately 79 per cent of
the group area (within the projected radius from the dominant 
central galaxy to the furthest group galaxy) was covered by the survey
footprint. Assuming galaxies follow a
Navarro, Frenk and White (NFW) radial density profile \citep{NFW96}
with a concentration 
half that of the dark matter and using the predicted mass from Table
\ref{TGROUPS}, this is equivalent to missing $\sim 23$ per cent of group
galaxies within the same radius. However, as we had full coverage of
the group core we did not exclude this group from our sample.

The number of FoF galaxies within each group is shown in Table \ref{TGROUPS}
for both the original G3Cv04 catalogue, on which our target list for
{\it Chandra} observations was based, and for the deeper G3Cv06 data\footnote{In addition to 
the increased depth of G3Cv06 ($r\leq19.8 ~\mathrm{mag}$ compared to $r\leq19.4 ~\mathrm{mag}$ for G3Cv04),
the updated catalogue also redetermined all galaxy redshifts with only a small
fraction of galaxies whose redshifts changed significantly. See \citet{LISKE15} for
further details}, which
has been utilised for the analysis presented in Section \ref{SXANALYSIS} onwards.
As one would expect, the multiplicity of most groups rises
somewhat with the deeper data. However, this is not guaranteed, since
the higher density of galaxies in G3Cv06 results in shorter linking lengths 
for the Friends-of-Friends analysis. As a result, some galaxies linked
to a group in the G3Cv04 catalogue may become disconnected in G3Cv06. The most
striking example of this is group 200130, which shows a 
substantial decrease in multiplicity from catalogue G3Cv04 to G3Cv06. Visual
inspection of this group within the G3Cv04 catalogue (ID 200003 in G3Cv04) indicates that there
is a substantial southern extension in the FoF group that is split
off as a separate group in G3Cv06 (ID 200477 in G3Cv06 with 16 members).

Observations of the 9 selected groups not already within the
\textit{Chandra} archive were completed by the ACIS-I instrument
on the \textit{Chandra} X-ray
Observatory in 2013. Observations of group 200130, an ACIS-S image
taken in 2002, were taken from the \textit{Chandra} archive.

\begin{table*}
\begin{minipage}{\textwidth}
\centering
	\caption{Summary of the predicted properties of our selected groups based on the G3Cv06 group catalogue and associated X-ray observations.}
	\label{TGROUPS}
	\begin{threeparttable}
	  \begin{tabular}{lccccccccccc}
	  \hline \hline
Group ID\tnote{a}	&	Central\tnote{b}	&	$\alpha$\tnote{c}	&	$\delta$\tnote{c}	&	$z$\tnote{c}	&	$N_{fof}$\tnote{d}	&	$M_{500, {\rm Pred}}$\tnote{e}	&	$r_{500, {\rm Pred}}$\tnote{e}	&	$T_{500, {\rm Pred}}$\tnote{f}	&	$t_{exp}$\tnote{g}	&	ObsID\tnote{g}	\\
			&	galaxy	&		&			&		&			&	($10^{13} ~h_{70}^{-1} ~\mathrm{M_{\sun}}$)	&	($h_{70}^{-1}$ kpc)			&	(keV)			&	(ks)		&		\\
\hline           
100053 (100072)	&	279874	&	139.74	&	1.149	&	0.0874	&	32 (23)	&	5.9	&	576	&	1.5	&	9.9	&	14001	\\
200015 (200011)	&	30699	&	176.53	&	-1.094	&	0.1175	&	34 (33)	&	4.8	&	532	&	1.4	&	15.8	&	14002	\\
200017 (200014)	&	536417	&	182.26	&	-0.965	&	0.1132	&	22 (20)	&	4.0	&	502	&	1.2	&	34.3	&	14005	\\
200043 (200018)	&	537303	&	184.71	&	-1.047	&	0.1195	&	23 (22)	&	5.2	&	548	&	1.4	&	10.6	&	14003	\\
200054 (200036)	&	136792	&	176.10	&	-1.851	&	0.1069	&	23 (17)	&	4.6	&	527	&	1.3	&	24.7	&	14004	\\
200099 (200034)	&	534758	&	174.93	&	-1.030	&	0.0777	&	23 (24)	&	4.7	&	536	&	1.3	&	14.9	&	14000	\\
200115 (200063)	&	136847	&	176.28	&	-1.758	&	0.0276	&	18 (17)	&	2.6	&	448	&	0.94	&	24.7	&	14007	\\
200130 (200003)	&	230776	&	181.11	&	1.896	&	0.0200	&	35 (46)	&	7.2	&	629	&	1.7	&	30.0	&	3234	\\
300008 (300006)	&	15899	&	217.19	&	0.708	&	0.1027	&	23 (20)	&	3.6	&	489	&	1.1	&	52.6	&	14006	\\
300067 (300028)	&	594961	&	222.75	&	-0.036	&	0.0429	&	22 (24)	&	2.6	&	446	&	0.94	&	25.7	&	14008	\\
\hline
    \end{tabular}
    \begin{tablenotes}
     \item[a] Group ID in parentheses shows the G3Cv04 group ID. Group IDs of the form $1xxxxx$, $2xxxxx$ and $3xxxxx$ indicate groups from GAMA regions G09, G12 and G15 respectively.
     \item[b] G3Cv06 galaxy ID for galaxies identified as optical central galaxies using our iterative centroid algorithm described in Section \ref{SCENTRE}. 
     \item[c] Listed centres correspond to the galaxy associated with the X-ray peaks. In cases were no emission is detected we assign a central galaxy as described in the Section \ref{SCENTRE}. For the group 200115 where the X-ray centroid is offset from the brightest group galaxies, we use the X-ray centroid. Redshifts are as defined by the \citet{ROBOTHAM11} FoF group finder.
     \item[d] Group FoF multiplicities from the G3Cv06 group catalogue (corresponding
values from the G3Cv04 catalogue are given in parentheses). 
     \item[e] $r$-band luminosity derived mass estimates as described in the main text. Predicted $r_{500}$ assumes the critical density of the Universe at redshift $z$ (Section \ref{SGRPSAMPLE}).
     \item[f] Temperatures estimated using the $M-T$ relation of \citet{SUN09}.
     \item[g] Exposure time and Chandra observation ID of each observation used. With the exception of the archival data for 200130, this was calculated using group properties from the G3Cv04 group catalogue.
    \end{tablenotes}
  \end{threeparttable}
\end{minipage}
\end{table*}

\subsection{Group Centres for X-ray Analysis} \label{SCENTRE}

Our X-ray analysis requires the location of an X-ray centre for each
group, about which we will extract spectral flux (or upper limits) and,
where possible, surface brightness profiles. Diffuse X-ray emission is observed in the majority of our sample,
and in most cases
the centroid of this emission is coincident with a bright group galaxy.
In these cases we adopt this galaxy as the centre of
the group. One group, 200115, has its X-ray emission centroid offset
from any bright galaxies. For this we adopt the X-ray 
centroid of the group as its centre.

Where there is no significant X-ray emission to help us locate the
bottom of the gravitational potential well, we use the G3Cv06 galaxy data
to define an optical group centre. For this we adopt a slightly
modified form of the iterative centring algorithm of
\citet{ROBOTHAM11}. The algorithm initially takes all member
galaxies and calculates a centroid weighted by galaxy luminosity. The
galaxy furthest from this centre is then removed from the sample and the
weighted centroid is recalculated. This procedure is iterated until only two
galaxies remain, the brightest of which is then identified as the central
galaxy.

We find that in a small number of cases this can be dominated by a bright
galaxy on the cluster outskirts. We modify this algorithm by assuming that central galaxies should be
near the centre of the velocity distribution. The weight of each
galaxy in the centroid calculation is therefore modified to include the inverse of
the line-of-sight velocity offset from the group mean, scaled by the
standard deviation of the velocity distribution,
$|z-\bar{z}|/\sigma_z$. 
At each iteration the mean velocity is recalculated,
whilst maintaining the standard deviation at its initial value. This
modification should prevent excessively bright galaxies in the cluster
outskirts dominating the weighted centroid of the group. We use this
algorithm to define a central galaxy for each group. In cases where an X-ray centroid is not possible, we use this optical
central galaxy as the group centroid. Ultimately this optical centre
was only required for group 200099, as explained in Section \ref{SXUNDETECT}.
The adopted centres of all 10 groups are listed in Table~\ref{TGROUPS}.

\section{X-ray Analysis} \label{SXANALYSIS}

In this section we discuss the reduction and analysis of our {\it
Chandra} data. The software packages \textsc{ciao} 4.5 and \textsc{Sherpa} (version 1 for \textsc{ciao} 4.5) were used for
the data reduction and the analysis of spectra and radial
profiles. The techniques used
were those of \citet{PASCUT14}, and we present only a brief outline here.

Our X-ray images were reduced from the initial level 1 event files
with the calibration files from \textsc{CalDB} 4.5.6. These
corrections include the effects of time-dependent gain variation and
charge transfer inefficiency. We additionally filtered out
flaring events by removing periods where the count rate was 20 per cent
greater than the median rate.

The resulting cleaned event files form the basis of our X-ray analysis. 
We first identify point sources within our X-ray images using \textsc{ciao}'s \textsc{wavdetect} tool. Once these are removed, the quality of our data 
allows us to detect diffuse X-ray emission in most of our target groups. 
In groups where we do not detect significant group-scale emission
we instead calculate
limits on X-ray properties as described in Section \ref{SXLIMITS}.

\subsection{Spectral Modelling} \label{SXSPEC}

We extract the diffuse X-ray emission in the energy range $0.3-3.0$ keV
within a radius of $0.5 r_{500}$ from our chosen
centre, excluding any point sources detected in the cleaned events
files. Initial estimates of $r_{500}$ are based on the GAMA luminosity-based 
mass estimates described previously. 
The extracted X-ray spectrum is fitted using a two-component
source + background model. The background is 
separately modelled, rather than being subtracted. This
allows the spectral fit to employ a maximum likelihood method using 
the Cash statistic \citep{CASH79}. This specifically allows for
the Poissonian nature of the data, and so is more appropriate
than $\chi^2$ for cases, such as most of ours, in which
the total number of photon counts is low \citep{HUMPHREY09}.

To establish an appropriate background model we 
extract counts from ACIS-I chips 0-3,
excluding point sources and the $0.5 r_{500}$ source region. An
additional \textsc{ciao} routine, \textsc{vtpdetect}, is used
to search for any other sources of diffuse emission in the field,
which are then also removed.
The cleaned background region is then fitted over the range of $0.3-7.0$
keV with a model which includes components for the cosmic soft X-ray and
particle background, galactic emission and instrumental lines
introduced by material along the optical path. We refer to
\citet{PASCUT14} for the specifics of this fit. In a small number of
cases an additional background term is required to
account for the effect of solar wind charge exchange.  In these cases
we model these with a set of Gaussians corresponding to the
O\textsc{vii}, O\textsc{viii}, Ne\textsc{ix} and Mg\textsc{xi} lines
at 0.56, 0.65, 0.91 and 1.34 keV respectively,
\citep[][]{KOUTROUMPA09, KUNTZ08}.

We then fit the source region as an APEC thermal plasma with the above background component.
We assume a fixed metalicity
of $Z=0.5Z_{\sun}$ \citep{SANDERSON09} relative to the \textsc{grsa}
cosmic abundance model \citep{GREVESSE98} and take the absorption column from the
galactic H\textsc{i} survey of \citet{KALBERLA05}, extracted using the
\textsc{nh} tool from the \textsc{HEASoft} software suite.

The source + background model is then fit. However, as the background itself consists
of two components, a vignetted photon background and a non-vignetted particle background,
simply rescaling the background to the source region would overcorrect the latter. We
therefore fit the source in two phases, first in the range $0.3-7.0$ keV to appropriately
rescale the particle background whilst providing an inital estimate of the source
properties. The background component is then fixed and the source re-fit to determine source
temperature, $T_{spec}$, and APEC normalisation, $\eta$, over the range $0.3-3.0$ keV. The APEC
model assumes $\eta = 10^{-14}/(4\pi[D_a(1+z)]^2) \int n_e n_H \mathrm{d}V$ where $D_a$ is the angular
diameter distance to the source at redshift $z$ in cm and $n_e$ and $n_H$ are the electron and hydrogen
number densities ($\mathrm{cm}^{-3}$) within a volume $\mathrm{d}V$.

Using the fitted temperature,
$T_{spec}$, we then revise our estimate of $r_{500}$ using the $r-T$
relation of \citet{SUN09}, \begin{equation}
 r_{500} = 602 h^{-1} \left(\frac{k_\mathrm{B}T_{spec}}{3~\mathrm{keV}}\right)^\frac{1.67}{3} ~\mathrm{kpc} \quad . 
\end{equation}

\noindent The spectrum is then re-extracted within the new
$0.5r_{500}$ aperture and refitted. If the newer estimate of $0.5r_{500}$
is larger than the initial value then we repeat the extraction and 
modelling the background, to avoid any possible contamination with
diffuse source flux. We adopt
the re-fitted $T_{spec}$ as our estimate of the system's mean
temperature and estimate group mass from this using the $M-T$
relation of \citet{SUN09}, \begin{equation}
 M_{500} = 1.27\times10^{14} h^{-1} \left(\frac{k_\mathrm{B}T_{spec}}{3~\mathrm{keV}}\right)^{1.67} ~\mathrm{M_{\sun}} \quad . 
\end{equation}

As shown by \citet{LEBRUN14}, the $M-T$ relation is relatively unaffected
by any feedback processes.  Therefore, masses estimated in this way should be
representative of the halo mass regardless of the thermal history of
the group (i.e. low or high entropy gas). In the case of groups with cool cores,
the mean gas temperature will differ somewhat depending on whether or not the
core region is excised when extracting the X-ray spectrum. Our data quality does 
not permit us to derive temperature profiles, so we do not attempt to excise core
emission. However the impact of cool cores on the global spectrum has been shown to
be small. \citet{RASMUSSEN07} show in their {\it Chandra} study that the central
temperature in groups drops only 10-20\% below the mean group temperature, even
in systems with strongly cooling cores (see their Figure 6), and \citet{OSMOND04} 
find that the impact of core excision on the mean spectral temperature of groups
is only $\approx$4\%. In practice, as we discuss in Section~\ref{SENTROPY} below,
with the exception of MKW4, our groups seem likely to contain at most very weak cool cores.

\subsection{Surface Brightness Profiles}

The surface brightness profiles of our groups are extracted by
binning the observed emission in annuli
centred on the positions listed in Table \ref{TGROUPS}. Bin widths are
chosen to give at least 15 counts per bin in our $0.3-3.0$ keV energy
band. We remove contaminating point sources and apply an exposure
correction. For groups without any strong cool core we assume that the
surface brightness profile can be represented by a single
$\beta$-model, \citep{CAVALIERE76}, \begin{equation}
 S(r) = S_0 \left(1+\left(\frac{r}{r_c}\right)^2\right)^{-3\beta +0.5} \quad , \label{EBETAPROF}
\end{equation}

\noindent where $r_c$ is the core radius, $\beta$ determines the slope
of the emission profile and $S_0$ is the central surface
brightness. In groups with a noticable excess of central emission we
modify the surface brightness model to be the sum of two
$\beta$-models, $S(r) = S_{core}(r) + S_{out}(r)$. This modification
is simpler than those used by, for example, \citet{ETTORI00} and
\citet{VIKHLININ06}. Due to the limited quality of the data from
most of our groups, fitting more complex models is unlikely to
provide significant improvements.

A flat background component, $S_{bg}$, is incorporated into the fitted surface
brightness model. This is not strictly correct, since the particle
background is not subject to vignetting, unlike the photon background, so
after exposure correction (which flattens the photon component of the
background) the particle contribution will actually rise with radius from
the optical axis of the instrument. However, as
most of our sources do not cover a very large fraction of the
ACIS-I field (radial extent typically less than or approximately a few hundred pixels
compared to a 2048 pixel wide field of view), and as our energy range
does not extend into the hard X-ray regime where particles
dominate the background, departures from a uniform background will
be small.

The surface brightness profiles are fitted using \textsc{Sherpa}. 
Due to limited statistics we do not fit all $\beta$-model components,
we fix $\beta_{out} = 0.5$, comparable to that observed for
low temperature groups \citep[e.g.][]{HELSDON00}, and $\beta_{core} =
2/3$, assuming that central emission has a standard slope comparable
to those measured in \citet{MOHR99}. 

\subsection{Luminosity}

To determine the bolometric X-ray luminosity of the diffuse group
emission we use the spectral fit
and the \textsc{Sherpa} algorithm \textsc{calc\_energy\_flux} in the
energy range 0.01 to 15 keV, applying the appropriate conversion from
flux to luminosity. As our extraction aperture is smaller than $r_{500}$ --
$0.5 r_{500}$ in most cases --
we estimate the luminosity, $L_{X,500}$, within $r_{500}$
by applying a rescaling based on the measured surface brightness profile.
The rescale factors typically range from 1.12 to 1.75. One 
group, 300067, requires
little rescaling (1.04) due to its centrally concentrated emission
profile, whilst the three groups 100053, 200099 and 200130, for which 
the extraction aperture is only 100 kpc (see Section \ref{SXLIMITS}),
have larger scale factors of 3-6. 

\subsection{Intracluster Medium Entropy}

We define entropy, $K$, as
\begin{equation}
 K(r) = \frac{k_\mathrm{B}T_{spec}}{n_e(r)^{2/3}} \quad , \label{EENTROPY}
\end{equation}

\noindent where we assume isothermal gas with temperature $T_{spec}$ and $n_e(r)$ is
the number density of electrons in the intragroup gas at radius $r$ \citep{VOIT05}. We note that the assumption
of an isothermal gas can lead to slight overestimates of central entropy in systems with cool cores.
To determine the gas
density follow the method of \citet{HUDSON10} to deproject $\beta$-model fits
to the surface brightness profile. For the single $\beta$-model this defines
\begin{equation}
 n_e(r) = n_{e,0}\left(1+\left(\frac{r}{r_c}\right)^2\right)^{-\frac{3\beta}{2}} \quad , \label{ENPROF}
\end{equation}

\noindent where $\beta$ and $r_c$ are derived from the surface brightness 
profile, Equation \eqref{EBETAPROF}, 
and $n_{e,0}$ is a normalisation factor calculated as
\begin{equation}
 n_{e,0} = \sqrt{\frac{n_{eH} 4 \pi D_a(z)^2 (1+z)^2 10^{14} \eta J }{EL}} \quad .
\end{equation}

\noindent Here $n_{eH}=1.176$ is the ratio of electron to hydrogen number densities within
a fully ionised plasma of $0.5Z_{\sun}$ metallicity (relative to the \textsc{grsa}
\citep{GREVESSE98} abundance tables) and $D_a(z)$ is the angular diameter distance to the group in cm 
at redshift $z$. $\eta$ is the normalisation of the APEC model fit in Section \ref{SXSPEC} and $EL$ is the ratio
of the emission integral within our extracted region to the central electron density. 

Equation \ref{ENPROF} can be extended to a double $\beta$-model fit as 
\begin{multline}
 n_e(r)^2 = n^2_{e,0,core}\left(1+\left(\frac{r}{r_{c, core}}\right)^2\right)^{-3\beta_{core}} \\
	  + n^2_{e,0,out}\left(1+\left(\frac{r}{r_{c, out}}\right)^2\right)^{-3\beta_{out}} \quad .
\end{multline}

\noindent where the labels `core' and `out' indicate the relevant property from the core and outer
$\beta$-model fit. We refer to \citet{HUDSON10} for the calculation of the central electron densities
$n_{e,0}$, $n_{e,0,core}$ and $n_{e,0,out}$.

Using the derived gas density profile and  Equation \eqref{EENTROPY} we
calculate the gas entropy at 
a radius of 10 kpc in each group to probe the core gas properties.

\subsection{Notes on individual groups}\label{SXLIMITS}

Some of the groups within the sample required modification to some aspects 
of our standard analysis procedure. These are described below.

\subsubsection{100053 and 200099}\label{SXUNDETECT}

Since their surface brightness profiles show no significant X-ray emission
on group scales, 
groups 100053 and 200099 are considered to be non-detections, and we instead
derive limits on the gas luminosity and entropy.

Examining the smoothed emission maps of these sources we find a small diffuse
source associated with a bright member galaxy for 100053 (see Figure \ref{FXOPT})
and adopt this galaxy as the centre of our analysis. 200099 appears featureless
in the smoothed images. We therefore adopt an  optical centre for this group based on
the iterative centroid algorithm described in Section \ref{SCENTRE}, though we note
that the centre identified by the algorithm in this case actually lies near
the edge of the group.

To determine limits on the gas properties for these two systems, we extract 
X-ray spectra within 100 kpc,
using a small, fixed aperture to increase the signal-to-noise relative to that 
within $0.5 r_{500}$. Assuming the $T_{pred}$ determined from the optical group
luminosity, we fit an APEC model with fixed temperature and metallicity. We then
adopt an upper limit corresponding to the $2 \sigma$ upper
bound on the fitted APEC normalisation  $\eta$. The
surface brightness profile is taken to be single $\beta$-model
with $\beta=0.5$ and $r_c = 0.2 r_{500}$, comparable to the mean 
of the other 8 groups in the sample ($0.17\pm0.03 ~r_{500}$). We assume $r_{500}$ 
as predicted by the luminosity mass estimate.

Using the $2\sigma$ upper limit on normalisation and the assumed surface brightness
profile, a $2\sigma$ upper limit on luminosity can be derived. Deprojecting this
surface brightness profile provides a $2\sigma$ upper limit on electron 
number density, $n_e$. Combining this with the assumed gas temperature, $T_{pred}$,
gives a $2\sigma$ \emph{lower} limit on central entropy. 

\subsubsection{200130}

Group 200130 is the known low redshift X-ray group MKW4 \citep{FUKAZAWA04}. This
group was imaged in 2002 with \textit{Chandra} in the ACIS-S configuration. We extract
our spectra from the back-illuminated S3 chip. However, due to the low redshift of 
this group the predicted $0.5 r_{500}$ aperture we would ordinarily use (313 kpc, 6.7 arcmin at $z = 0.02$), 
extends beyond the chip boundary. Additionally, this system is known to have
traceable emission across the ACIS-S CCD \citep[][]{SUN09}, rendering our usual
approach of measuring a local background unsuitable.

We instead use the Markevitch blank sky background datasets to estimate the background \citep[][]{VIKHLININ05,SUN09}.
Using the S3 chip only we extract a spectrum within a 100 kpc aperture, comparable
to the size of the chip. We do this for both the background and data. We scale the
background to match the hard X-ray counts and subtract this from our source 
spectrum. We then fit with a source model using a $\chi^2$ statistic.

To determine the radial profile, as we do not have data beyond 100 kpc, we again
make use of the scaled blank sky background to constrain background emission. 
We use the core radius and slope of the outer gas halo,
$r_c = 204$ kpc and $\beta = 0.64$, determined by \citet{VIKHLININ99}
using ROSAT imaging data that extended to much larger radius. 
We perform a two $\beta$-profile fit of the background
subtracted radial profile where we fit only the amplitude of the outer $\beta$
model but allow the inner, core profile freedom to optimise normalisation, 
core radius and slope.

We calculate luminosity as described previously, extrapolating from 100 kpc 
to $r_{500}$. Profiles of temperature for this group are available from the literature
\citep[e.g.][]{VIKHLININ06, SUN09}. However for consistency with the treatment
of our other groups, we adopt a single mean temperature, $T_{spec}$, calculated
from an emission weighted temperature profile. We use the temperature profile from 
\citet{VIKHLININ06} and, weighting by the surface brightness profile, average
within $0.5 r_{500}$, adopting $r_{500}=538 ~\mathrm{kpc}$ initially \citep{SUN09}. 
This temperature is then used to recalculate $r_{500}(T)$, and the mean
temperature is recomputed, iterating until convergence.
The final mean temperature is used for our entropy estimate.

\subsubsection{200115}

Group 200115 features a diffuse X-ray source not associated with any member
galaxy. There are no background groups or clusters in the GAMA survey
that this emission may be associated with, nor are any known groups or
clusters within $1\arcmin$ of the emission found within the NASA Extragalactic 
Database\footnote{The NASA/IPAC Extragalactic Database (NED) is operated by the
Jet Propulsion Laboratory, California Institute of Technology, under contract
with the National Aeronautics and Space Administration.}. 
We therefore attribute the emission to the group, and 
estimate its centroid in the 0.3--3.0 keV band within a 100 kpc 
aperture (at $z=0.028$), finding this to be offset by 130 kpc from 
the central galaxy identified by our iterative centre-of-light algorithm. 
Another bright galaxy, only $0.05 ~\mathrm{Mag}$ fainter, is offset by 32 kpc
from the X-ray centroid. We note that such offsets have been seen
in other groups and clusters which appear to have been subject 
to recent disturbance. We use the X-ray centroid as the centre for our 
subsequent analysis.

\subsubsection{200054}

Fitting the spectrum of group 200054 over our standard 0.3--3.0 keV band we 
find a high temperature of $5.2_{-1.9}^{+4.9} ~\mathrm{keV}$. This motivates 
us to extend the upper energy band, whereupon the fitted temperature
drops substantially to $\sim 3 ~\mathrm{keV}$. 
We therefore opt to fit this group within the full $0.3-7.0 ~\mathrm{keV}$
band used when rescaling the background. We additionally take $\beta_{out} = 2/3$,
consistent with high temperature systems, rather than the lower value,
$\beta_{out}=0.5$, used for the cooler groups within this work.

\subsubsection{300067}

The emission within this group was observed to be highly centrally concentrated. Attempting to
fit the two $\beta$-model to the surface brightness profile found a negligible contribution
from the second, outer $\beta$-model. We instead fit a single profile.

\begin{table*}
\begin{minipage}{\textwidth}
\centering
	\caption{Results of the X-ray analysis for our sample of optically selected groups.}
	\label{TRESULTS}
	\begin{threeparttable}
	  \begin{tabular}{lcccccc}
	  \hline \hline
GroupID	&	$T_{spec}$\tnote{a}	&	$L_{X,500}$\tnote{b}		&	$M_{500}$\tnote{c}	&	$r_{500}$\tnote{c}	&	$K_{10\mathrm{kpc}}$	&	$f_{500, gas}$\tnote{d}	\\
	&	(keV)			&	($10^{42} \mathrm{erg ~s^{-1}}$)	&	($10^{13} M_{\sun}$)	&	(kpc)			&	($\mathrm{keV} ~\mathrm{cm}^2$)	&	\\
\hline
100053	&	$(1.53)$    		&	$<1.70$		&	(5.91)		&	(576)		&	$>267$		&	$<0.026$	\\
200015	&	$1.05_{-0.12}^{+0.21}$	&	$3.02\pm0.66$	&	$3.16\pm0.83$	&	$480\pm42$	&	$52.9\pm13.7$	&	$0.045\pm0.007$	\\
200017	&	$1.34_{-0.23}^{+0.56}$	&	$1.36\pm0.46$	&	$4.74\pm2.33$	&	$550\pm90$	&	$52.7\pm11.9$	&	$0.025\pm0.003$	\\
200043	&	$0.97_{-0.25}^{+0.37}$	&	$1.63\pm0.62$	&	$2.77\pm1.47$	&	$460\pm81$	&	$45.1\pm20.6$	&	$0.034\pm0.008$	\\
200054	&	$2.80_{-0.53}^{+1.03}$	&	$12.5\pm1.61$	&	$16.2\pm7.53$	&	$828\pm129$	&	$110\pm80.0$	&	$0.037\pm0.010$	\\
200099	&	$(1.34)$		&	$<1.24$		&	(4.71)		&	(536)		&	$>243$		&	$<0.025$	\\
200115	&	$0.59_{-0.10}^{+0.09}$	&	$0.37\pm0.06$	&	$1.20\pm0.12$	&	$347\pm30$	&	$64.5\pm20.6$	&	$0.024\pm0.004$	\\
200130	&	$1.79_{-0.02}^{+0.02}$	&	$27.2\pm0.6$	&	$7.66\pm0.15$	&	$645\pm4$	&	$26.9\pm0.4$	&	$0.075\pm0.002$	\\
300008	&	$1.67_{-0.23}^{+0.31}$	&	$1.99\pm0.30$	&	$6.80\pm1.83$	&	$620\pm56$	&	$97.7\pm38.4$	&	$0.023\pm0.004$	\\
300067	&	$0.90_{-0.08}^{+0.11}$	&	$0.39\pm0.06$	&	$2.44\pm0.44$	&	$440\pm27$	&	$35.8\pm4.9$	&	$0.008\pm0.001$	\\
\hline
    \end{tabular}
     \begin{tablenotes}
	\item[] Values in parentheses are derived from GAMA mass and temperature estimates
	\item[a] Mean temperature within an aperture of $\approx 0.5 r_{500}$.
	\item[b] Extrapolated using the surface brightness fits from the extraction aperture to $r_{500}$.
	\item[c] Derived from the \citet{SUN09} mass -- temperature and radius -- temperature relations for groups and clusters.
	\item[d] Estimated gas mass fractions within $r_{500}$ (Section \ref{SGASMASS}).
     \end{tablenotes}
    \end{threeparttable}
\end{minipage}
\end{table*}

\section{Results} \label{SRESULTS}

In this section we present the results for our sample of optically selected
galaxy groups. The results of the spectral fitting and derived properties are shown in Table \ref{TRESULTS}
whilst Table \ref{TRESULTSSB} presents the results of the surface brightness fits to
each group. Diffuse X-ray emission is detected in 8 of our 10 groups, a larger fraction
than detected in studies such as the XI project \citep[only 1 bright source
and 2 weak detections out of 9 targets, ][]{RASMUSSEN06} where 
substructure/virialisation was not considered when selecting groups. 
Our detected fraction
is consistent with that found in the optically selected groups studied 
by \citet{BALOGH11}, who found 5 groups were undetected in a sample of 18 targets.
Interestingly, their sample, which was drawn from the 2PIGG catalogue \citep{EKE04}
with a narrow mass range ($3\times10^{14} < M/\mathrm{M_{\sun}}<6\times10^{14}$),
was additionally selected to exclude groups with non-Gaussian velocity distributions.
Whilst these samples are small, the difference between the XI result and that 
presented here, does indicate that selection by substructure can substantially improve
the reliability of a group sample.

In Figure \ref{FXOPT} we show the optical
SDSS images with X-ray contours overlaid. Of the 8 groups where we detect X-ray 
emission, there are three (200015, 200054 and 200115) where the central galaxy 
identified by our iterative centre of light algorithm
(Section \ref{SCENTRE}) is not associated with the peak of the X-ray emission.

\begin{table}
\centering
	\caption{X-ray surface brightness profiles.}
	\label{TRESULTSSB}
	\begin{threeparttable}
	  \begin{tabular}{lcccc}
	  \hline \hline
GroupID	&	$r_{c,core}$\tnote{a}	&	$\beta_{core}$\tnote{a}	&	$r_{c,out}$\tnote{a}	&	$\beta_{out}$\tnote{a}	\\
	&	(kpc)		&			&	(kpc)		&			\\
\hline
100053	&	-		&	-	&	(115)	&	(0.5)	\\
200015	&	$18\pm17$	&	(0.66)	&	$107\pm31$	&	(0.5)	\\
200017	&	$10\pm6$	&	(0.66)	&	$77\pm28$	&	(0.5)	\\
200043	&	$14\pm23$	&	(0.66)	&	$87\pm48$	&	(0.5)	\\
200054	&	$25\pm33$	&	(0.66)	&	$160\pm24$	&	(0.66)	\\
200099	&	-		&	-	&	(107)		&	(0.5)	\\
200115	&	-		&	-	&	$36\pm49$	&	(0.5)	\\
200130	&	$4.0\pm0.8$	&	$0.444\pm0.004$	&	(204)\tnote{b}	&	(0.64)\tnote{b}	\\
300008	&	$9.4\pm19.1$	&	(0.5)	&	$88\pm15$	&	(0.66)	\\
300067	&	$15\pm4$	&	(0.66)	&	-		&	-	\\
\hline
    \end{tabular}
     \begin{tablenotes}
	\item[a] Surface brightness profiles assuming the double $\beta$-model described in \S\ref{SXANALYSIS}. In cases of non-detection or where only one $\beta$-model is sufficient we report only one set of model parameters. Values in parentheses are fixed as described in the text and not allowed to fit.
	\item[b] Outer surface brightness profile parameters derived by \citet{VIKHLININ99} from a fit to ROSAT PSPC data.
     \end{tablenotes}
    \end{threeparttable}

\end{table}

\begin{figure*}
\begin{minipage}{\textwidth}
  \centering
      \parbox{150mm}{
	\includegraphics[width=75mm]{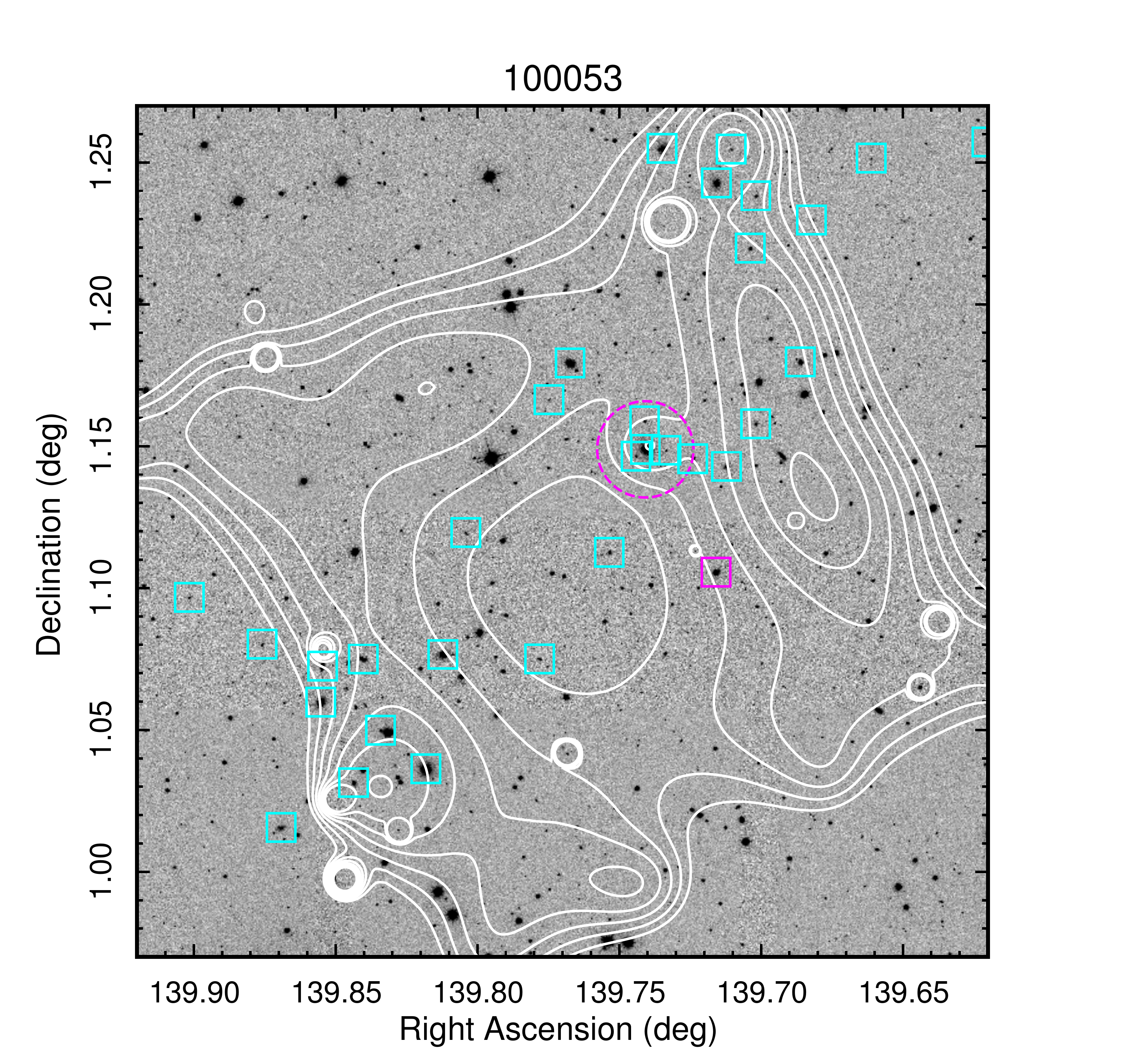}
	\includegraphics[width=75mm]{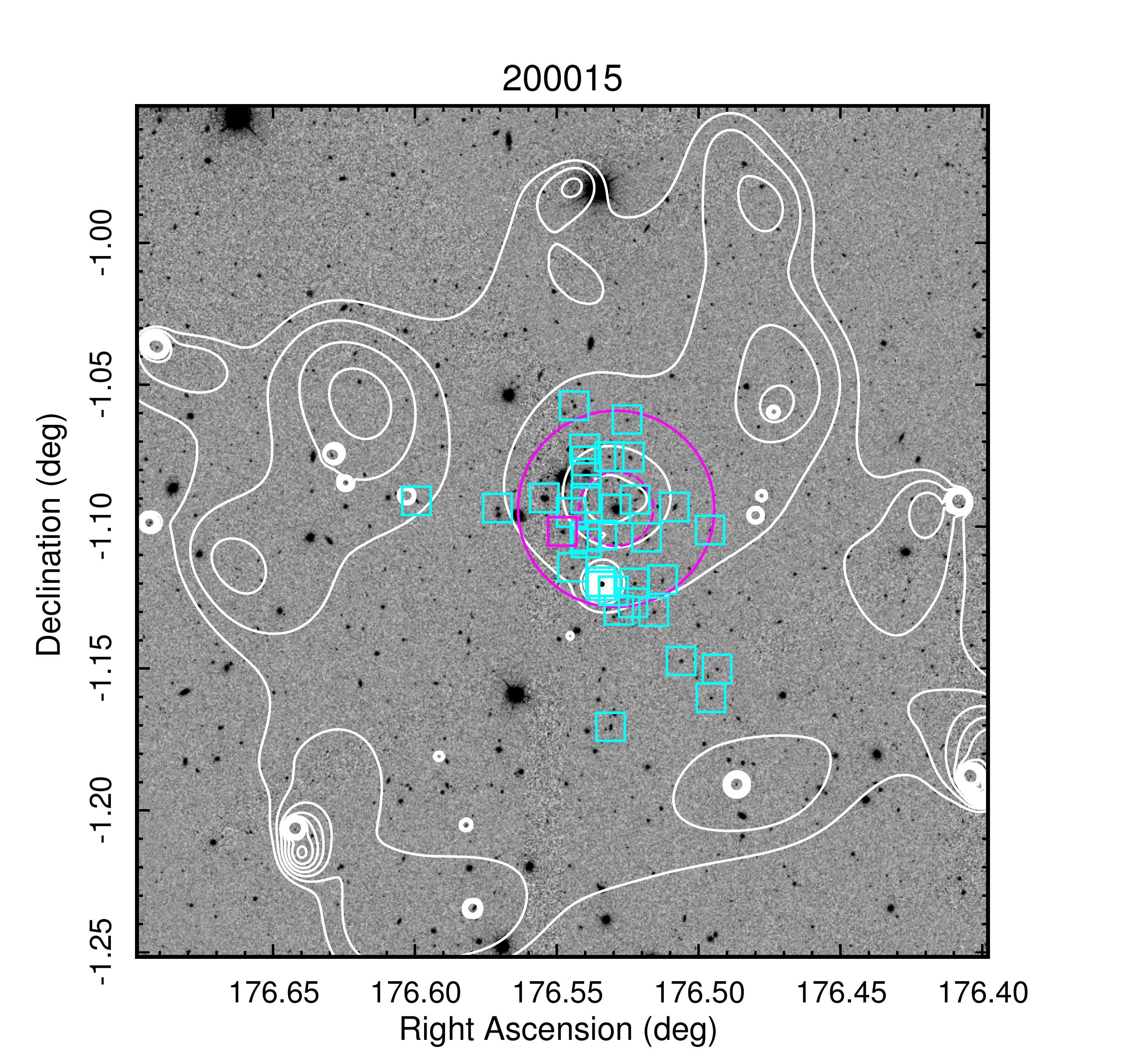}
	\\
	\includegraphics[width=75mm]{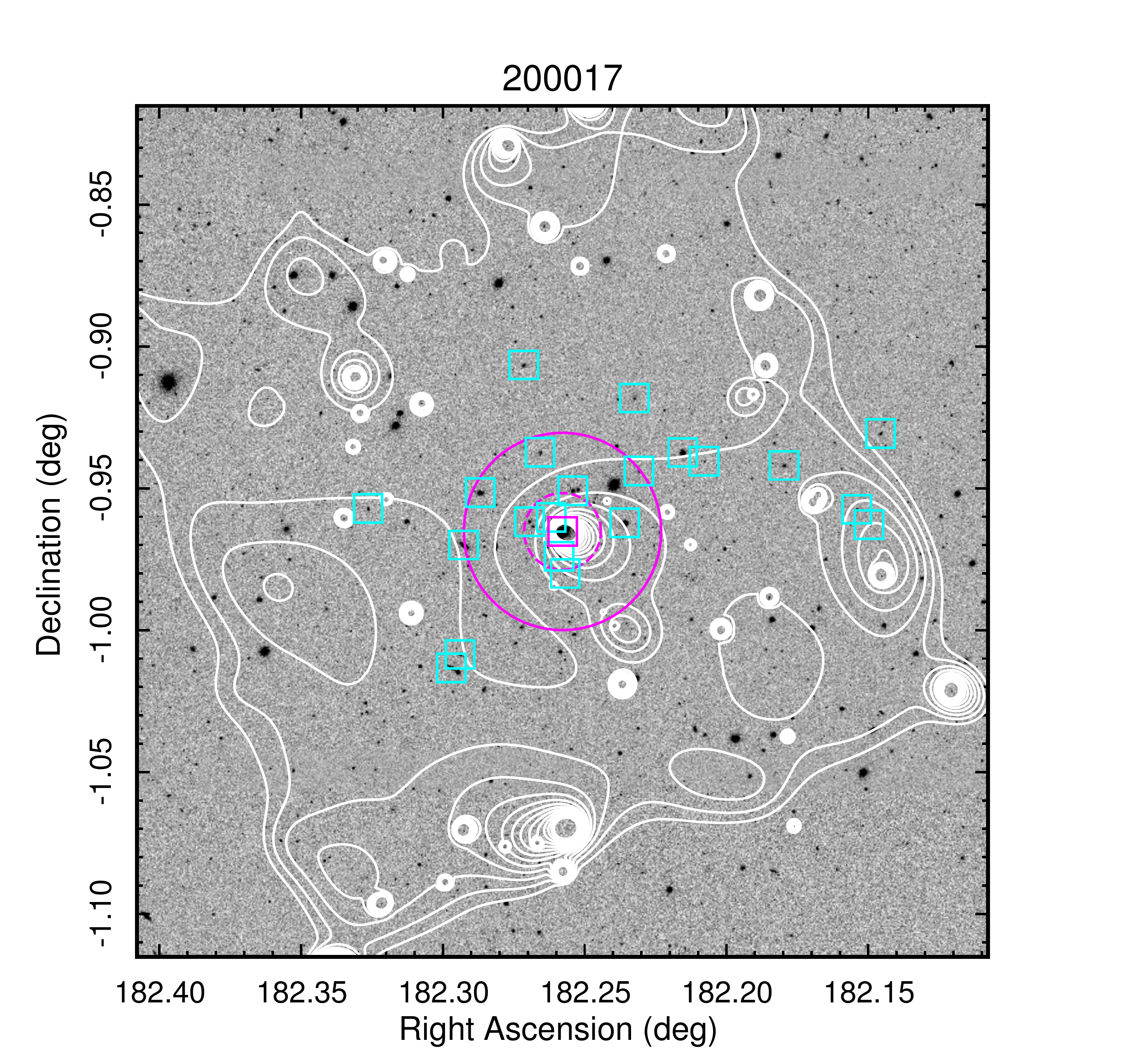}
	\includegraphics[width=75mm]{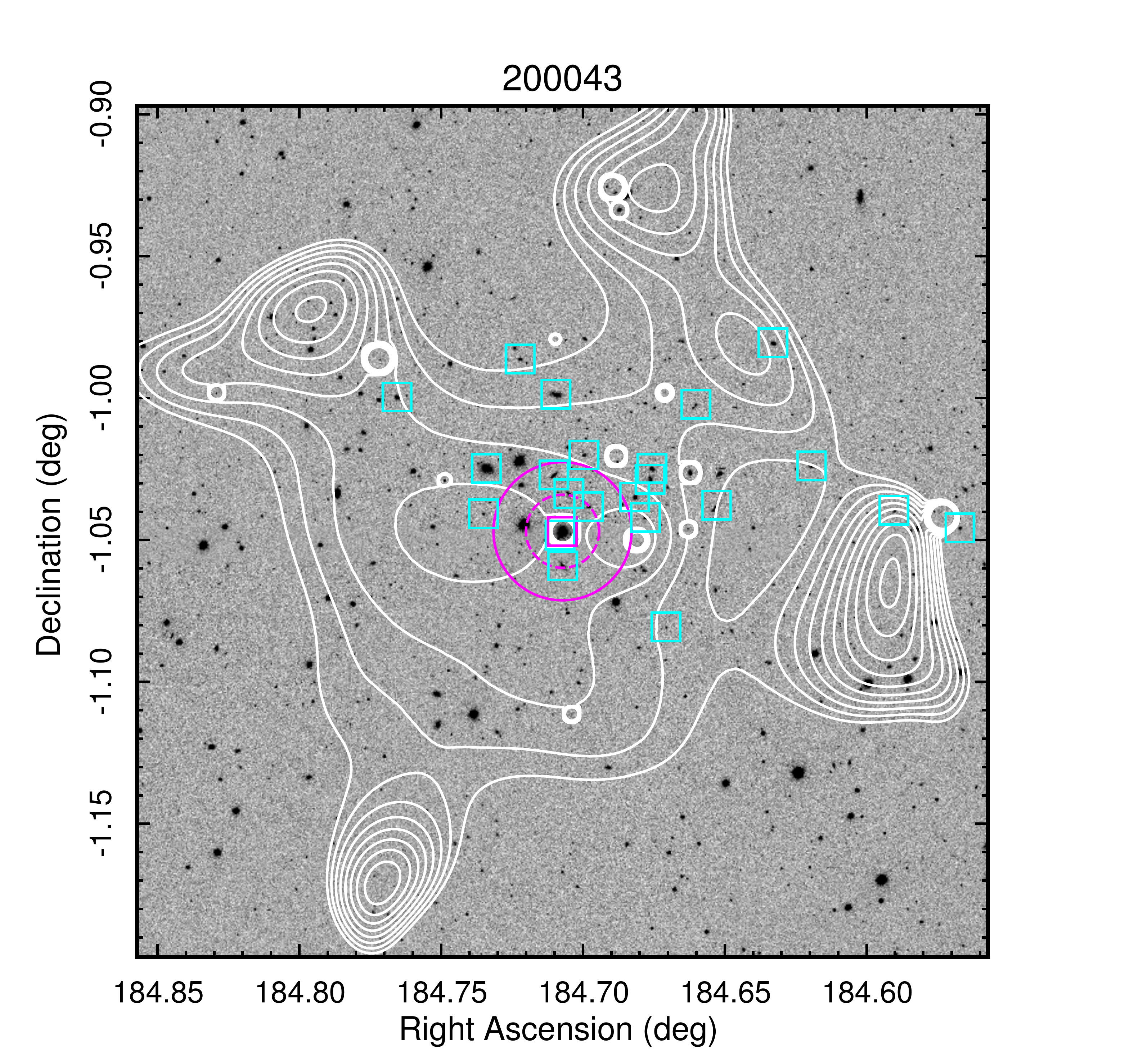}
	}
	\caption{Optical images of our groups from the SDSS with X-ray contours overlaid (\textit{white lines}). The X-ray contours are derived from adaptively smoothed images of the observed X-ray emission in our analysis band (0.3--3.0 keV) set at arbitrary levels for illustration purposes only. Also shown are the G3Cv06 member galaxies (\textit{cyan squares}) with central galaxies (as defined by our iterative centre of light algorithm defined in Section \ref{SCENTRE}) marked by \textit{magenta squares}. For scale we illustrate a 100 kpc region with a \textit{dashed magenta circle} and the extraction region, when different, by the \textit{solid magenta circle} centred on the X-ray source to be analysed (either the X-ray centroid or an optical centre, Section \ref{SCENTRE}).}
	\label{FXOPT}
\end{minipage}
\end{figure*}

\begin{figure*}
\ContinuedFloat
\begin{minipage}{\textwidth}
  \centering
      \parbox{150mm}{
	\includegraphics[width=75mm]{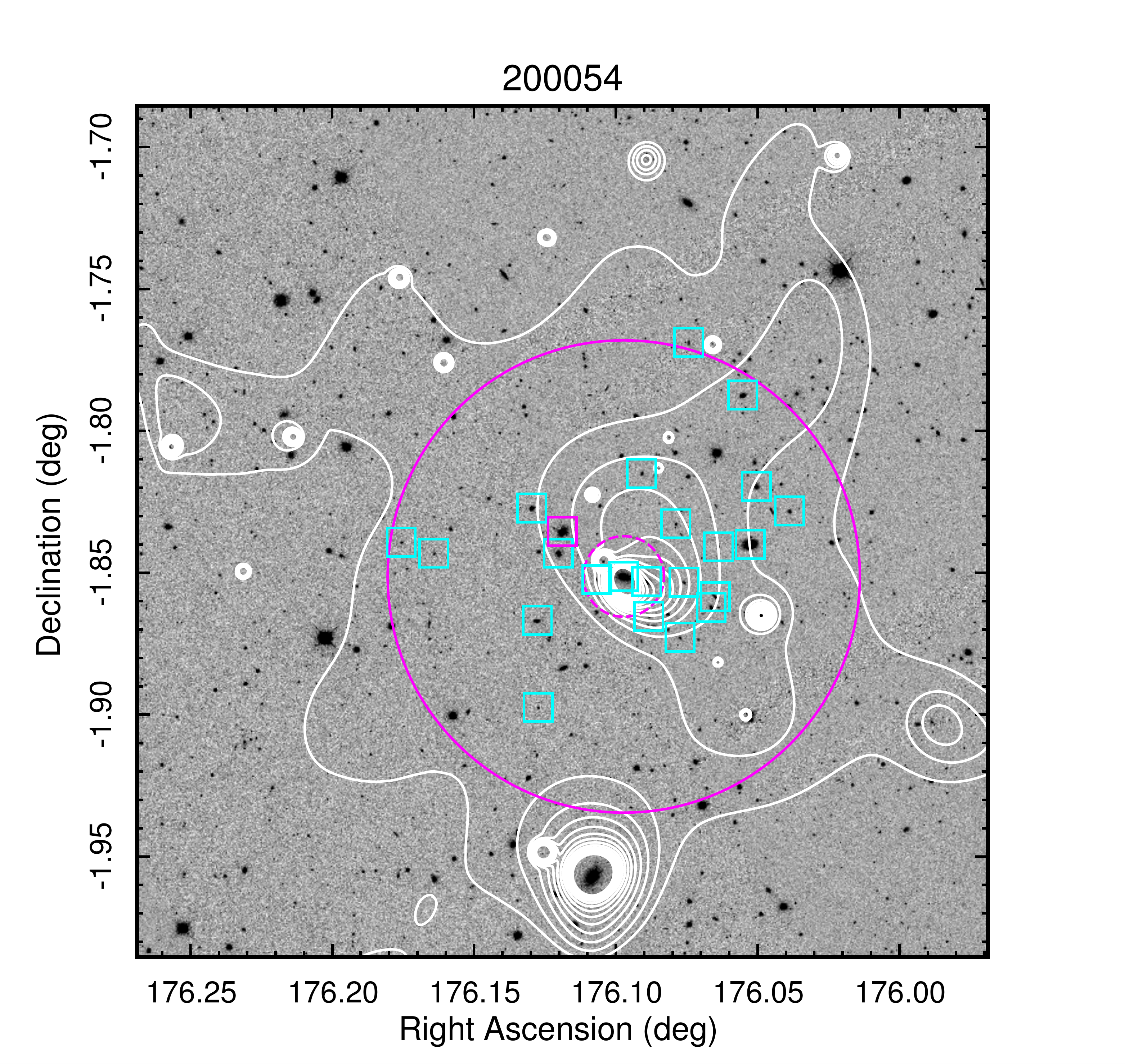}
	\includegraphics[width=75mm]{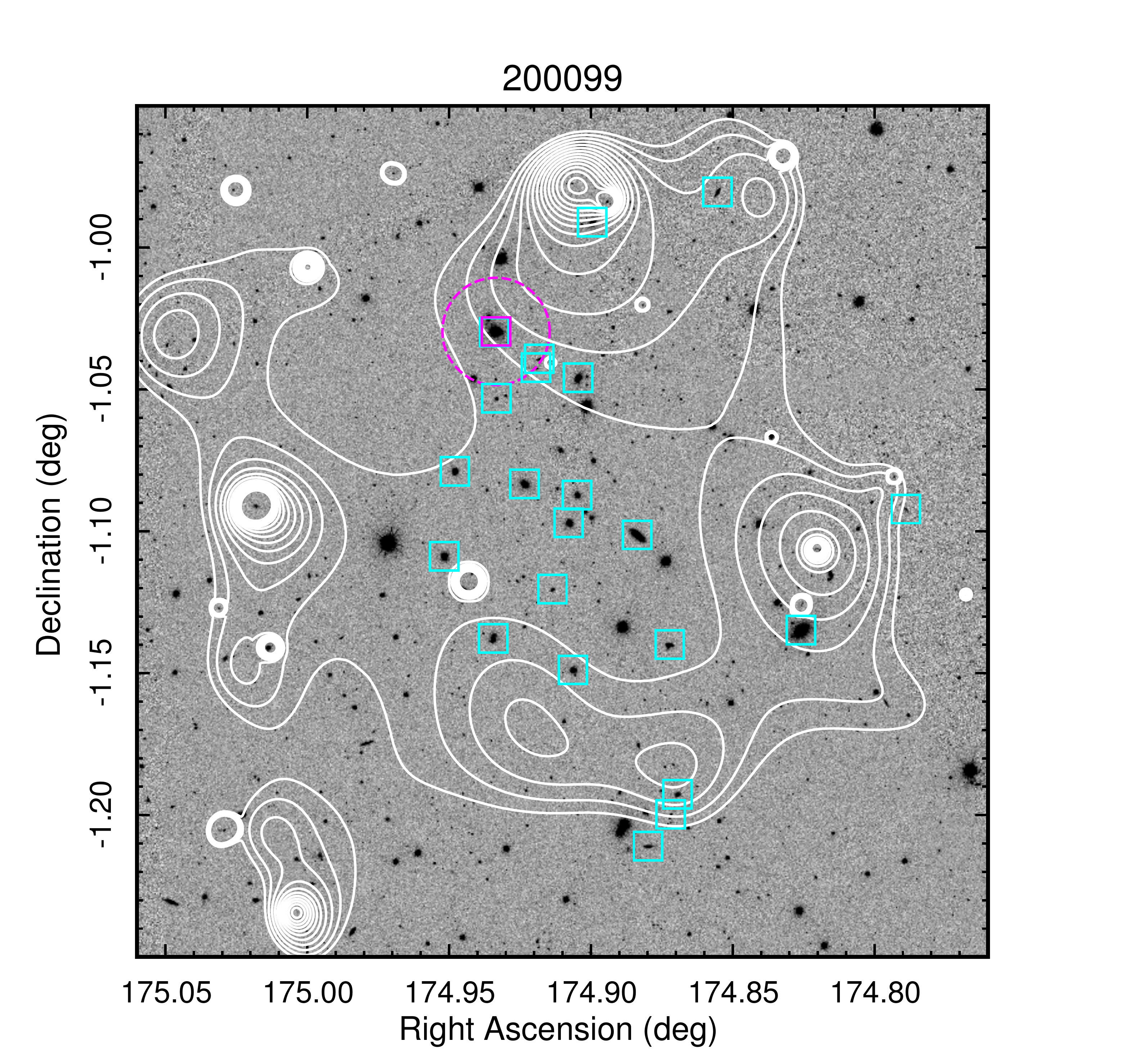}
	\\
	\includegraphics[width=75mm]{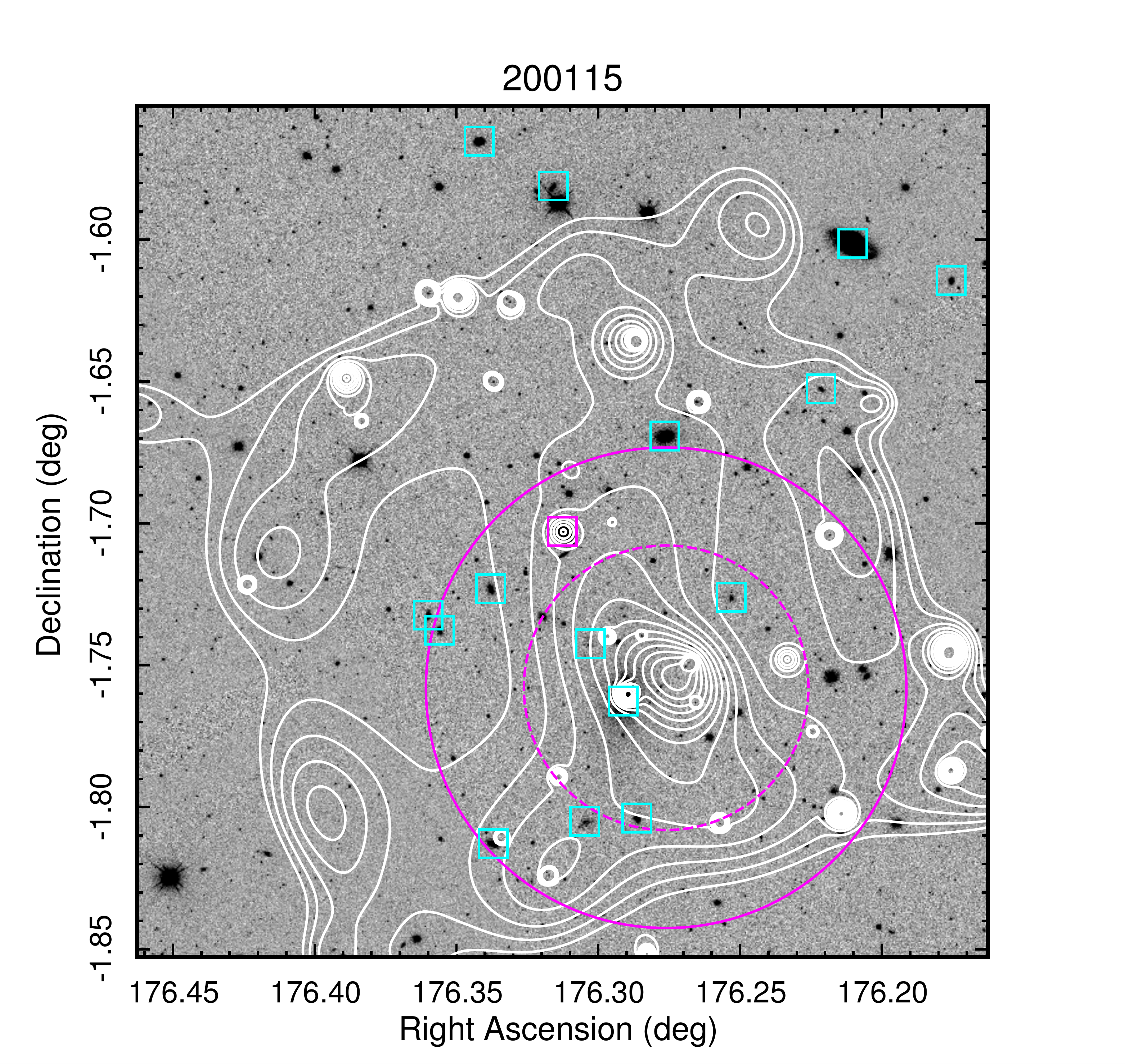}
	\includegraphics[width=75mm]{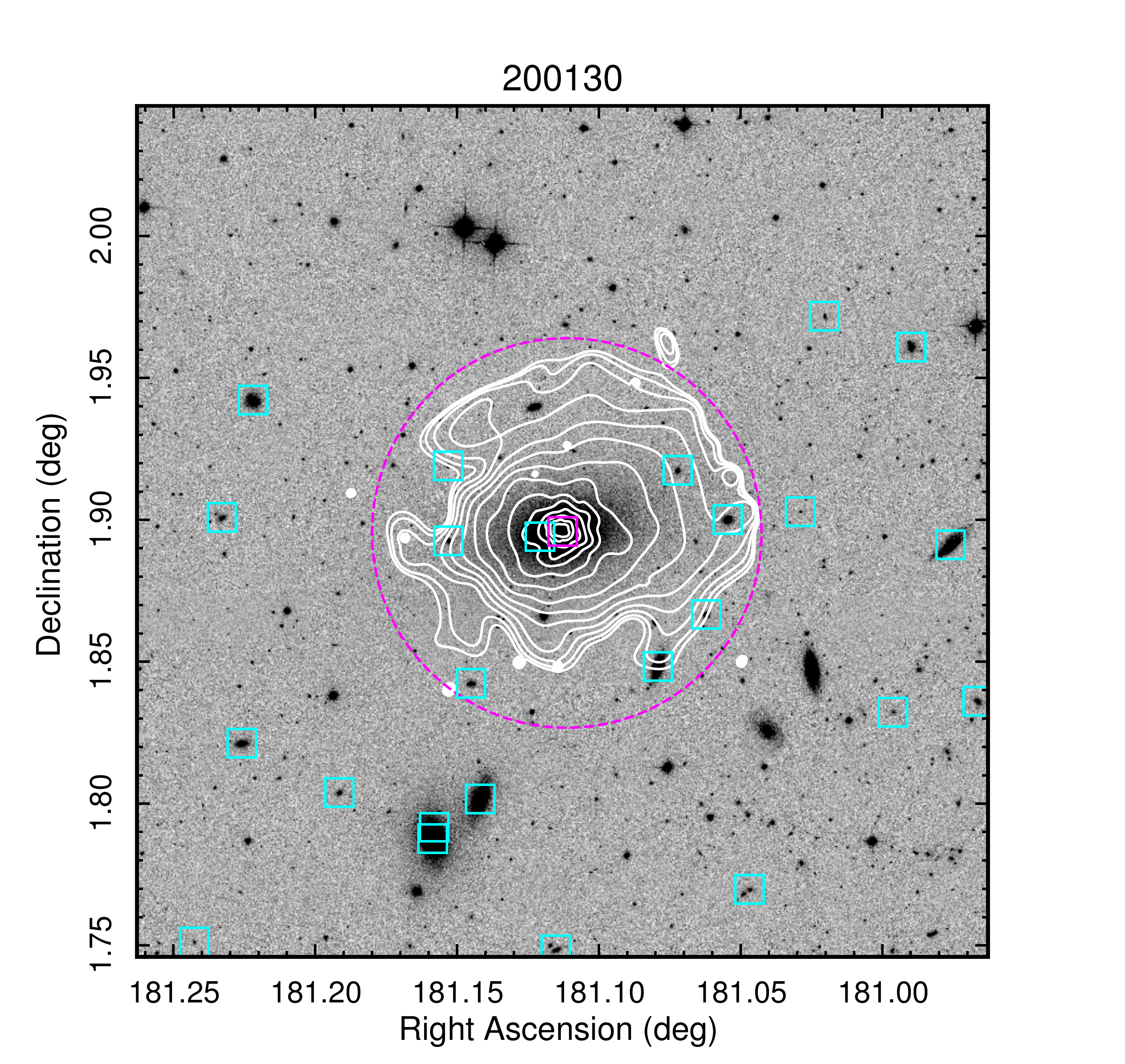}
	\\
	\includegraphics[width=75mm]{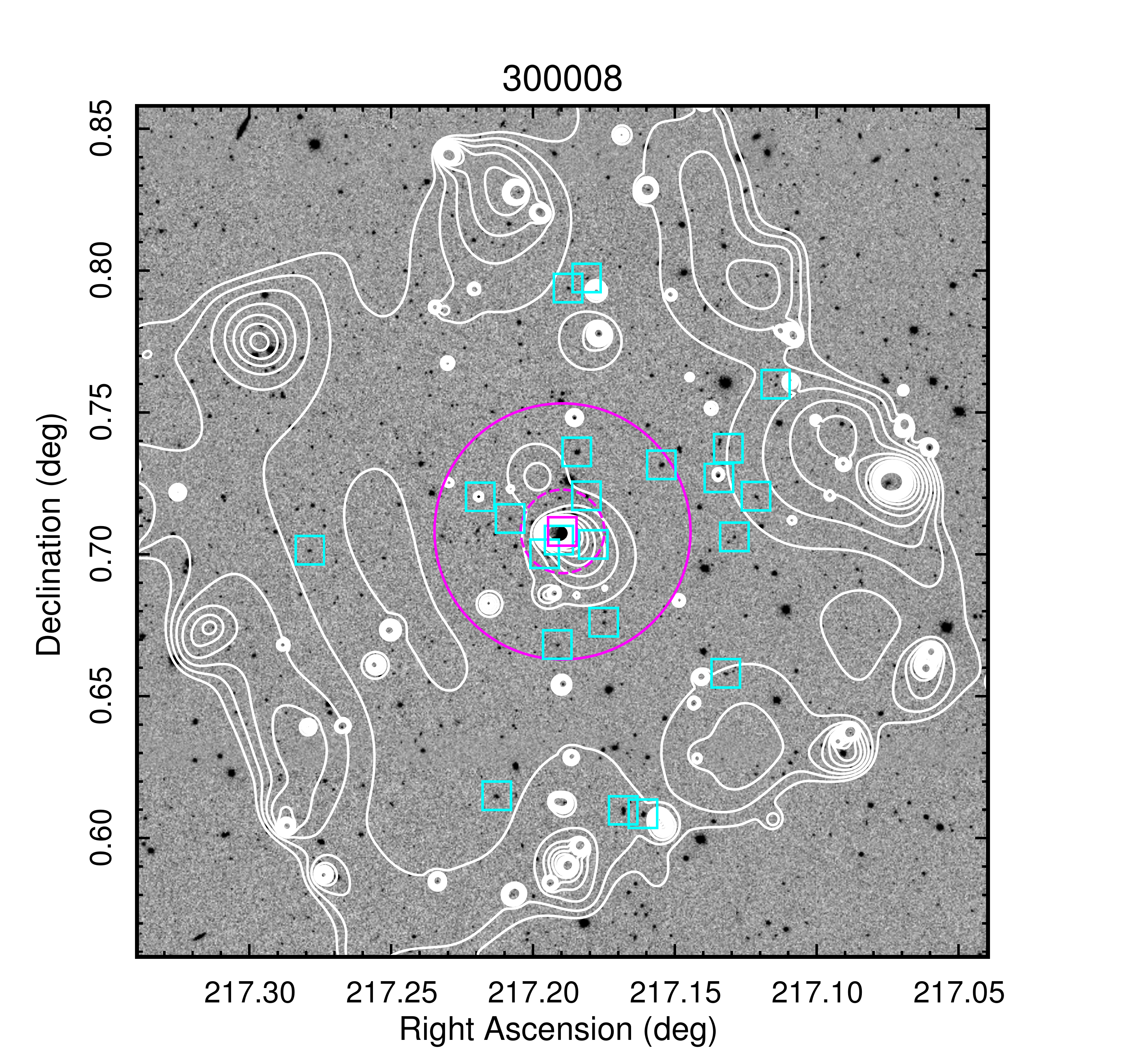}
	\includegraphics[width=75mm]{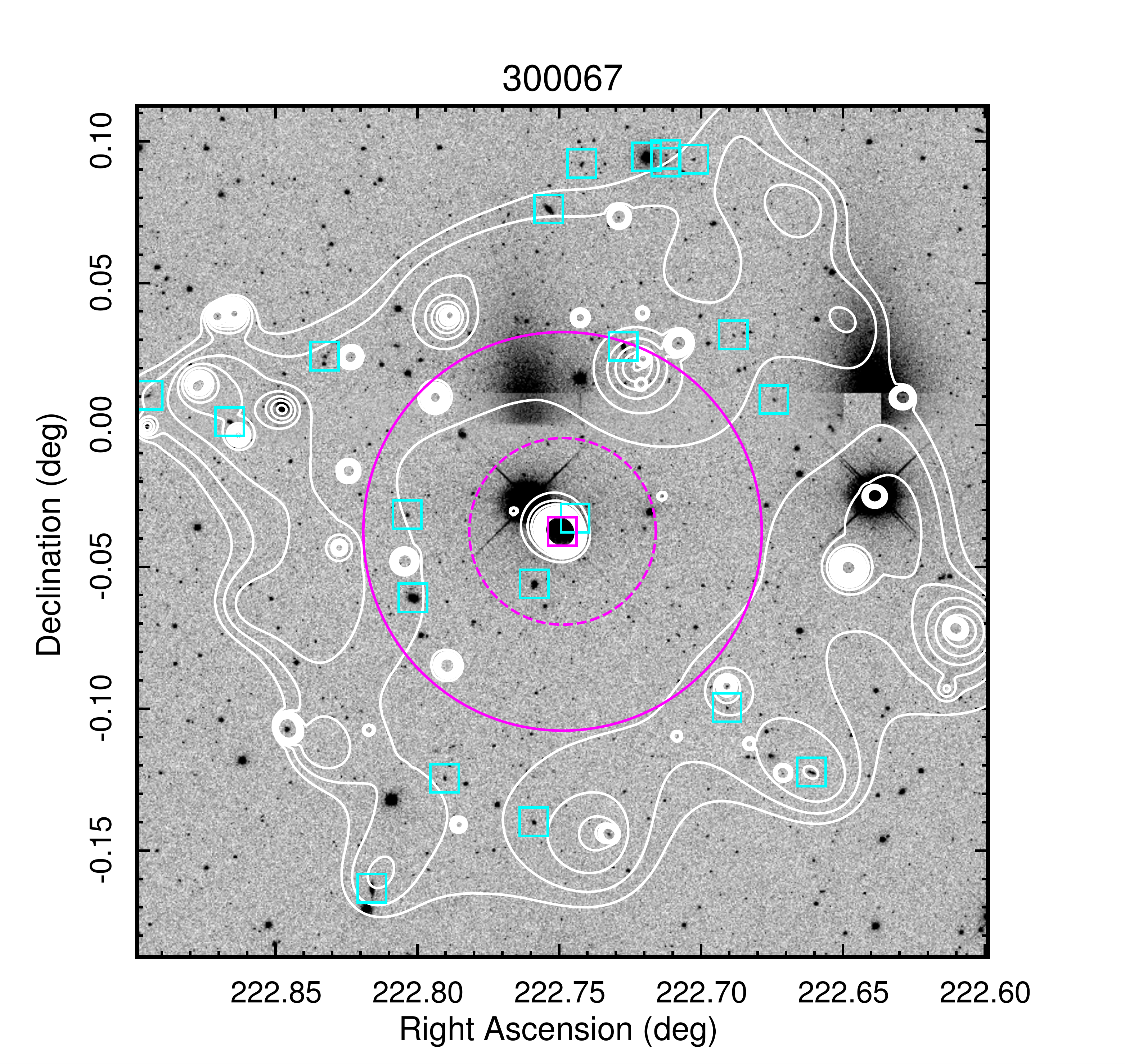}
	}
	\caption{(Continued)}
	\label{FXOPT2}
\end{minipage}
\end{figure*}

\subsection{$L_X - T_{spec}$}\label{SRESLT}

An initial assessment of our sample can be made through their position on the
X-ray luminosity-temperature relation ($L-T$). This relation, which has a self-similar 
expectation of $L_X \propto T^2$ for systems with temperatures above $\sim 3$ keV,
and a flatter slope at lower temperatures \citep[$L_X \propto T$, ][]{BALOGH99} where line emission is significant,
has been shown to be significantly steeper, especially in the group regime
 \citep[e.g.][]{HELSDON00, OSMOND04, PRATT09, SLACK14},
with slopes of 3 to 4. This similarity breaking is attributed to feedback processes
that inject entropy into the gas halo \citep[e.g. ][]{BALOGH99, BABUL02, VOIT01}, suppressing X-ray
luminosity in gas cores. 

\begin{figure}
\centering
	\includegraphics[width=0.49\textwidth]{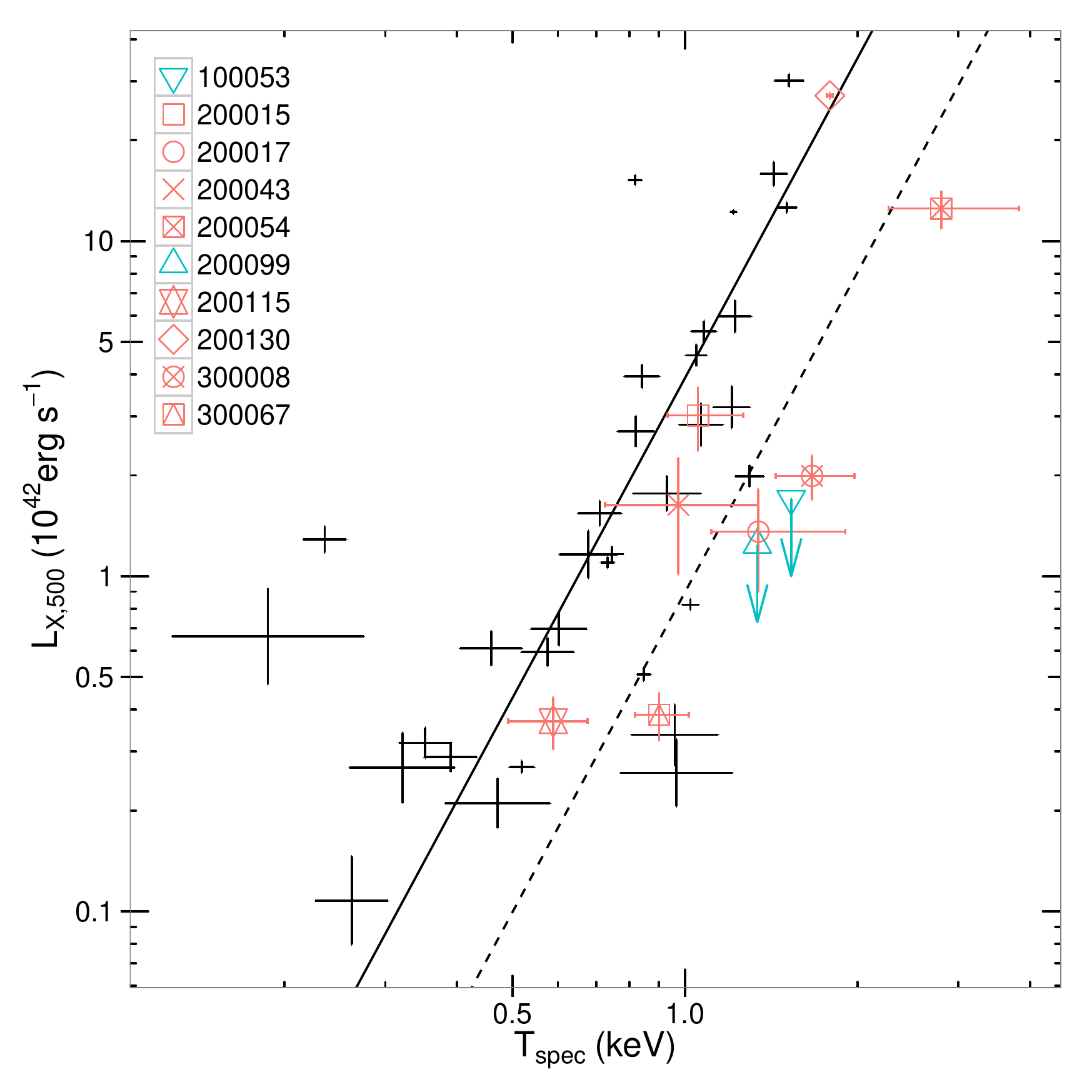}
	\caption[Luminosity -- temperature relation.]{The luminosity -- temperature relation for our group sample (\textit{coloured points}). Overlaid are groups from the \citet{OSMOND04} sample (\textit{black points}). The \textit{solid line} represents the $L - T$ relation of \citet{SLACK14}. We show the $L-T$ relation of our sample, modified from the \citet{SLACK14} relation by refitting the normalisation only, excluding the undetected groups 100053 and 200099, as the \textit{dashed line}.}
	\label{FLT}
\end{figure}

In Figure \ref{FLT} we show our group sample overplotted on the $L-T$ group
data from the GEMS sample of \citet{OSMOND04}. Also shown is the $L-T$ relation (for groups only) 
found by \citet{SLACK14} using a compilation of several group and cluster studies
spanning nearly 2 orders of magnitude in temperature,
\begin{equation}
  L_{X,500} = 1.27\times10^{44} \left(\frac{k_\mathrm{B}T}{3~\mathrm{keV}}\right)^{3.17} \mathrm{erg ~s^{-1}} \quad .
\end{equation}

With the exception of the archival group MKW4 (200130), we find that
our groups have X-ray luminosities below those observed for the GEMS
sample and expected from the $L-T$ relation for typical X-ray group
samples.  Quantifying the
size of our luminosity decrement relative to the literature $L-T$
relation by fitting the normalisation of the relation to our
data (excluding the two non-detections), we find that our optically
selected sample is underluminous by a factor $4$ relative to an X-ray
selected one. This renormalisation is shown as the dashed line on
Figure \ref{FLT}. This deficit is in qualitative agreement with the
results of \citet{ANDERSON14} who find optical groups in a stacked
analysis to be a factor 2 underluminous on the $L-M$ relation.

Despite the group luminosities being low compared to the standard
relation, our measured group temperatures show no
significant bias relative to temperatures predicted from the 
optical luminosity-based mass estimates using the \citet{SUN09} $M-T$ relation. 
Excluding the two non-detections, and MKW4 whose
analysis differed to that of the other groups, the mean offset
between predicted and observed temperatures
($\log_{10}(T_{pred}) - \log_{10}(T_{spec})$) is only $-0.005\pm0.072$, increasing
slightly to $-0.007\pm0.062$ if MKW4 is included. This suggests that
the mass estimation used here is on average unbiased, though
with substantial scatter.  This scatter, totalling 0.18 dex, has $0.09$
dex contributed by the measurement error on $T_{spec}$, with the
remaining 0.15 dex introduced through uncertainty on the calibration
of the optical luminosity mass estimation and intrinsic scatter on
that relation.

\subsection{Entropy} \label{SENTROPY}

\begin{figure}
\centering
	\includegraphics[width=0.49\textwidth]{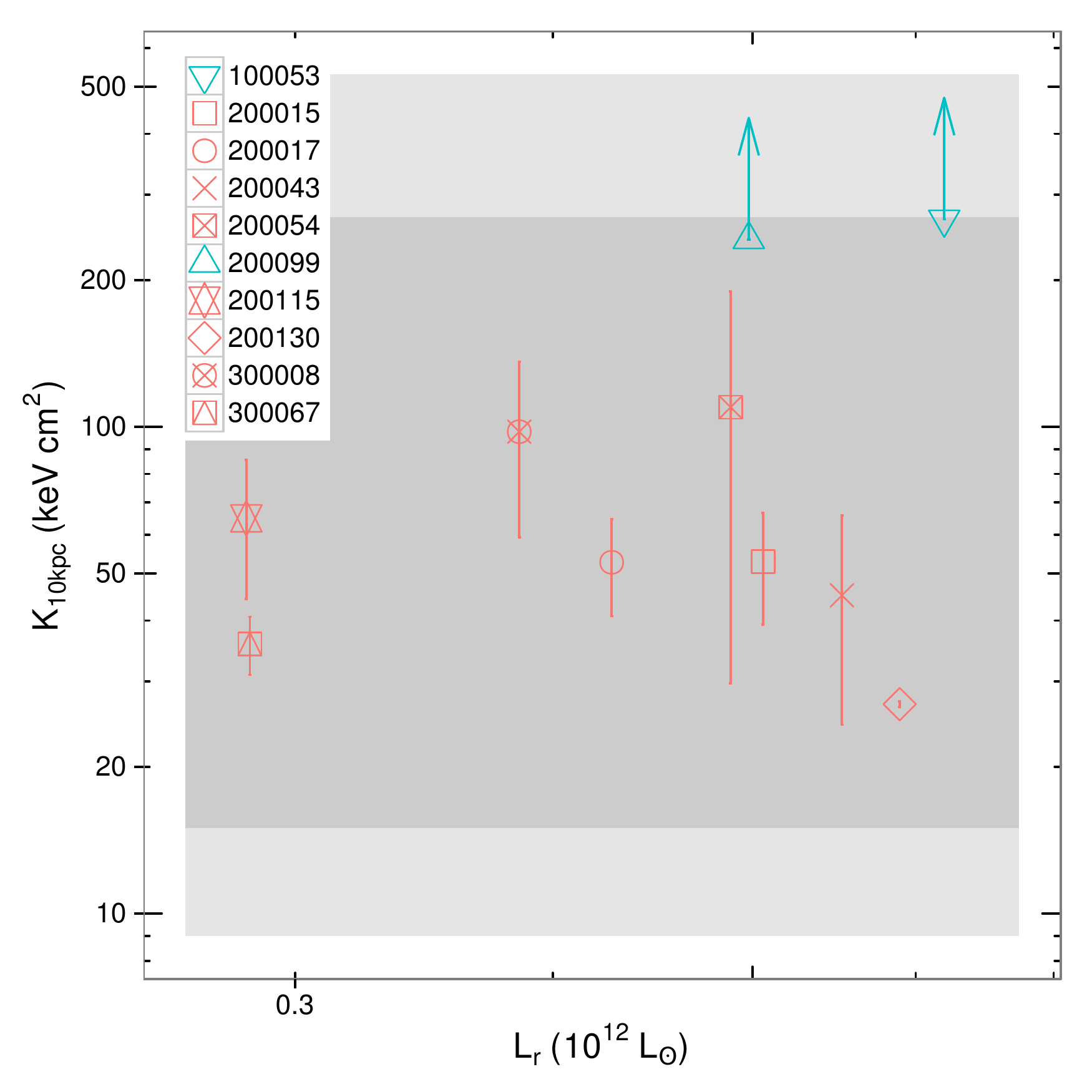}
	\caption{Central (10 kpc) entropies against total optical $r$-band luminosity. We show the $1\sigma$ (\textit{dark shading}) and $2\sigma$ (\textit{light shading}) upper and lower limits at 10 kpc derived from the OWLS simulations \citep{MCCARTHY11}.}
	\label{FKLOPT}
\end{figure}

In Figure \ref{FKLOPT} we plot central entropies, calculated from
Equation \ref{EENTROPY} using the surface brightness parameters in
Table \ref{TRESULTSSB} to estimate gas density, against optical
luminosity. We plot against $L_r$ to avoid any correlation that may
be introduced by plotting against an X-ray derived quantity. Note that the
group sample spans a fairly narrow range in $L_r$, as a
result of the selection on richness and predicted {\it Chandra} exposure
time. This is not a problem for our study, since we are effectively
exploring the full range in gas properties for groups over a limited mass
range. In the OWLS project simulations, the feedback model that best reproduces the 
observed entropy distribution, as well as other halo baryon properties such as
stellar mass fractions, incorporates both AGN and supernova feedback, together 
with radiative cooling \citep{MCCARTHY11}. The predicted range in central
entropies from this model is shown in Figure \ref{FKLOPT} by the
shaded regions. The majority of our groups lie well within the
$1\sigma$ range, however we note that our two non-detections have $2
\sigma$ lower limits substantially higher than the central entropies
of the rest of the sample. These two groups are plausible candidates for 
members of the population of high entropy groups predicted by the strong feedback models.

\begin{figure}
\centering
	\includegraphics[width=0.49\textwidth]{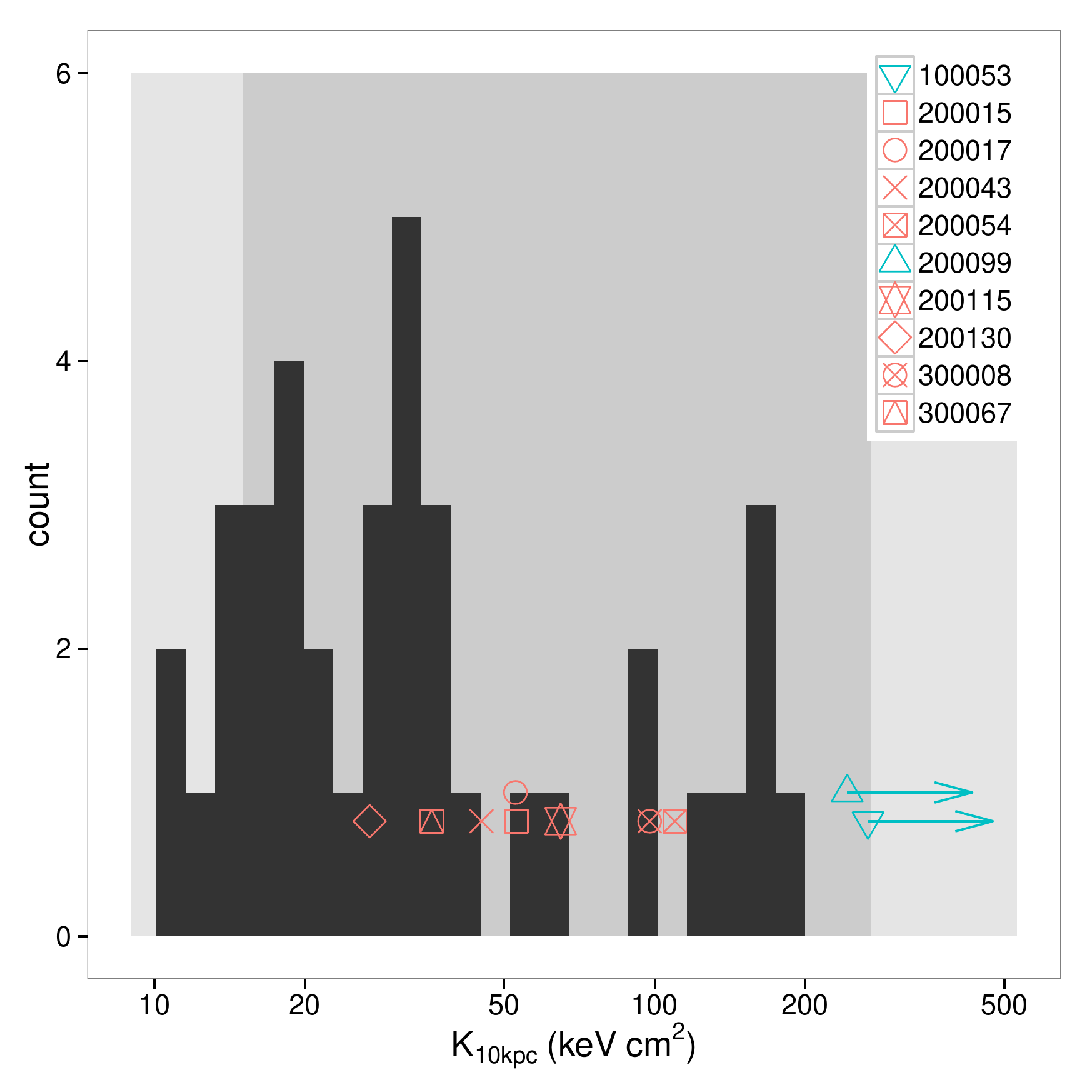}
	\caption{Histogram of group entropies at 10 Kpc using the entropy profiles of the ACCEPT sample \citep{CAVAGNOLO09} with $k_\mathrm{B}T<3~\mathrm{keV}$ with our measured entropies overlaid. As in Figure \ref{FKLOPT} we show the $1\sigma$ (\textit{dark shading}) and $2\sigma$ (\textit{light shading}) upper and lower limits at 10 kpc derived from the OWLS simulations \citep{MCCARTHY11}.}
	\label{FKLOPTCAVAGNOLO}
\end{figure}

In Figure \ref{FKLOPTCAVAGNOLO}
we compare these results to the distribution of central entropies
for groups and clusters within the ACCEPT \citep{CAVAGNOLO09}
data base\footnote{\url{http://www.pa.msu.edu/astro/MC2/accept/}}. We
use entropy profiles $K(r) = K_0 +
K_{100}(r/100\mathrm{kpc})^\alpha$ with their fitted values of $K_0,
K_{100}$ and $\alpha$ to estimate group entropies at 10 kpc. For
comparability with our sample we have taken only groups with $T_X < 3
~\mathrm{keV}$, which reduces the ACCEPT sample from 241 systems to 
38. The estimated entropies at 10 kpc for our X-ray detected systems mostly lie above the main peak in the entropy distribution for the 
\citet{CAVAGNOLO09} sample, and a two-sided Kolmogorov-Smirnov test gives a low probability (p=0.04) that the two distributions are consistent. The sense of this disagreement, whereby the majority of
our entropy estimates appear elevated relative to the general ACCEPT
population, is consistent with the lower X-ray luminosity observed. Moreover, the high entropy
group candidates, 100053 and 200099, have lower limits on entropy which are higher than the largest central entropies seen in the cool ACCEPT sample.

Since MKW4 (our group 200130) is included in ACCEPT, we can compare
the value of entropy in this derived from our own analysis, with that
calculated from the \citet{CAVAGNOLO09} data.
As shown in Table \ref{TRESULTS} we determine a central
entropy of $26.9 \pm 0.4 ~\mathrm{keV ~cm^2}$, compared to the value of $28.4
~\mathrm{keV ~cm^2}$ derived from the ACCEPT profile. Whilst the
ACCEPT value lies outside our small 1$\sigma$ confidence region, 
this small error bar results from the high
statistical quality of the data for this particular system. (As can be seen 
from the Table, entropy errors for all other groups in our sample are
at least an order of magnitude larger.) In practice, much larger discrepancies
can be expected to follow from the simplicity of our method (the isothermality
assumption, for example, compared to the temperature
deprojection used for ACCEPT), so we regard agreement to $\sim$5 per cent as 
quite satisfactory.

To estimate the lower limits on group entropy for groups 100053 and
200099, a series of assumptions needed to be made regarding the
temperature and gas density profile. We investigate these assumptions
and the impact deviations from them might have on the estimated limits
in Section \ref{SDISCUSSION}.

As entropy and cooling times are inherently linked properties of group
gas (at the same temperature high gas entropy implies low gas density
and therefore a reduced cooling rate), we can use our measured
entropies to explore the cool core (CC) / non-cool core (NCC)
status of the groups in our sample. Our choice of the threshold separating these 
two populations is guided by the observed bimodality in central entropies
reported by \citet{CAVAGNOLO09}, which split the population at
30-50~$\mathrm{keV ~cm^2}$. Groups below this threshold were found to be more
likely to show features associated with active cooling
\citep{CAVAGNOLO08}. Whilst the \citet{CAVAGNOLO09} split is
based on the value of the central baseline value
($K_0$) derived from fitting the $K(r)$ model discussed above,
our radius of 10 kpc should be small enough for the entropy
to have dropped to close to the baseline value, so it is reasonable
to adopt a similar threshold value.
Accordingly we classify groups with $K_{10\mathrm{kpc}}<30~\mathrm{keV ~cm^2}$ as 
containing probable cool cores.

From Table \ref{TRESULTS} we see that only one group falls below this threshold and is
therefore likely to be a CC group. Though two of our groups are plausibly moderate NCC groups,
with central entropies lying just above or consistent within $1\sigma$ of the CC threshold,
the remainder of our sample, though consistent with the CC threshold at a $2\sigma$ level, are likely NCC groups. Cooling times can be estimated
as $t_{cool} \propto K^{3/2} T^{-1/2} \Lambda(T)^{-1}$, 
where $\Lambda(T)$ is the cooling function.
For pure bremsstrahlung emission
where $\Lambda(T) \propto T^{1/2}$ \citep[e.g.][]{DONAHUE06}
this results in $t_{cool} \propto K^{3/2} T^{-1}$. At low
temperatures where line emission becomes significant, such as in the groups
observed here, the temperature dependence of $\Lambda(T)$, and hence
of $t_{cool}(T)$ flattens \citep[e.g. ][]{MCKEE77, BALOGH99}.  
With the exception of group 200130 which has the second highest temperature, three of our four
lowest entropy groups are also amongst the coolest
groups in this sample and should therefore have the lowest cooling times.

We note in the case of group 300067 the diffuse X-ray emission is confined 
to a small region around a central galaxy. This is reminiscent of the compact galaxy coronae
observed by \citet{SUN07}. Its radial extent ($\sim 15$ kpc) is larger than 
those seen by \citet{SUN07} ($\sim 4$ kpc),  but its temperature is comparable.
The \citet{SUN07} coronae were found around galaxies
within hot cluster environments. This clearly differs to the
environment seen in 300067, where no other diffuse emission is
observed. It is possible that this compact X-ray halo is surrounded by undetectable
high entropy gas, but we do not label this group as a high entropy
candidate.

\subsection{Gas Mass Fraction} \label{SGASMASS}

\begin{figure}
\centering
	\includegraphics[width=0.49\textwidth]{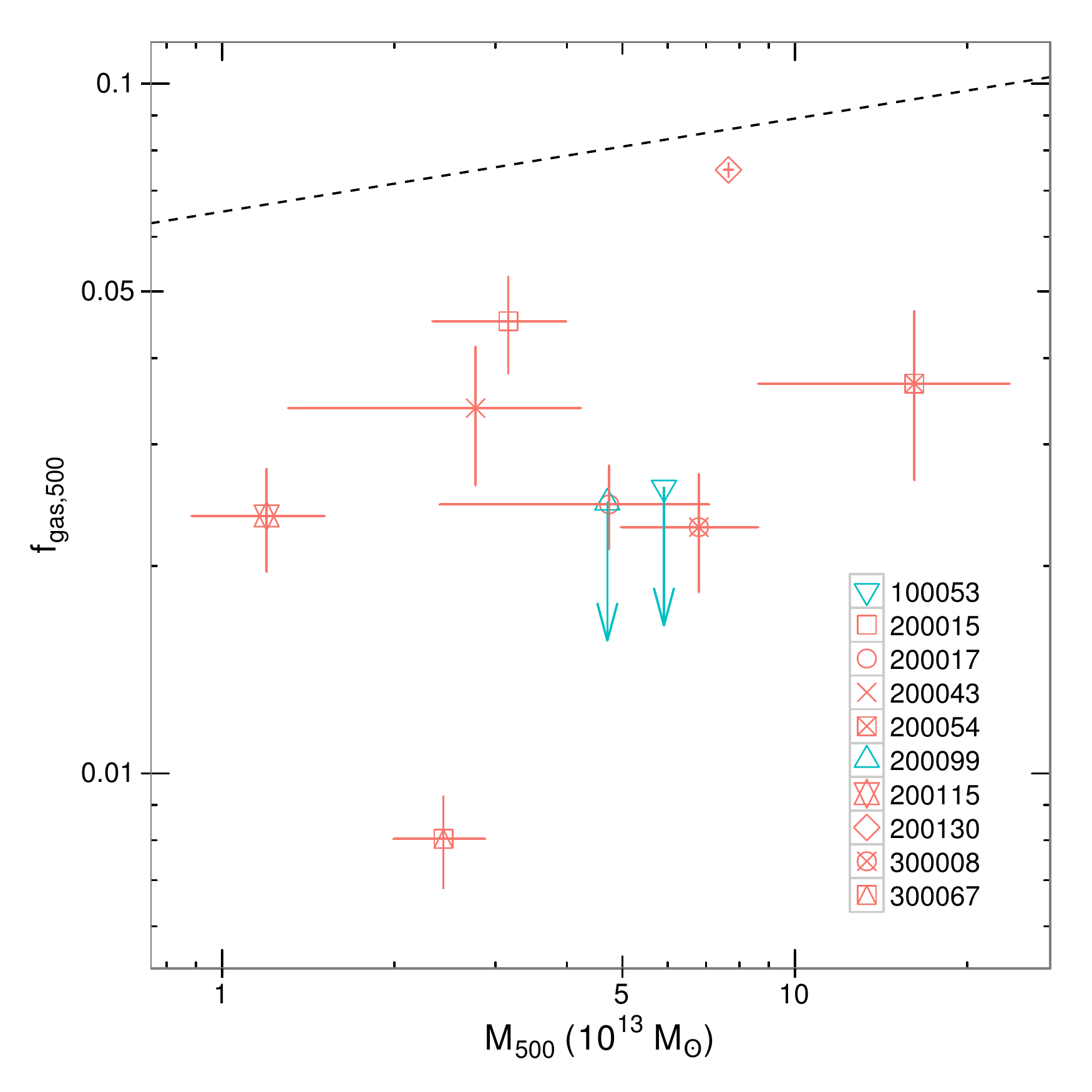}
	\caption{Gas mass fraction within $r_{500}$ for our 10 groups. Colour code and key as per Fig. \ref{FLT}, with \textit{blue} points
	indicating the groups for which only a $2 \sigma$ upper limit on gas mass has been determined. For comparison, the dashed line
	indicates the best fit $f^{\rm gas}_{500} - M_{500}$ relation found by \citet{SUN09} for their sample of X-ray selected groups and clusters.}
	\label{FMGAS}
\end{figure}

Using the gas number density profiles determined above, $n(r)$, we can also determine the gas mass fractions
for each group. We follow the method of \citet{SANDERSON13} such that
\begin{equation}
 M_{500}^{\rm gas} = m_{e} \int_0^{r_{500}} 4\pi r^2 n_e(r) ~\mathrm{d}r \quad ,
\end{equation}

\noindent where the factor $m_{e} = 1.159 ~\textrm{amu}$ is the gas mass per electron ($1 ~\mathrm{amu} = 1.66 \times 10^{-27}~\mathrm{kg}$)
for a fully ionised plasma of $0.5Z_{\sun}$ metallicity (again, relative to the \textsc{grsa} \citep{GREVESSE98} abundance tables).
For groups 100053 and 200099 we determine a $2 \sigma$ upper limit on gas
mass using the standard $\beta$ model and parameters estimated from
GAMA optical data and limits from the X-ray data. The resulting upper limits
on the gas mass fractions are shown in Figure \ref{FMGAS}
and in Table \ref{TRESULTS}.

As might be expected from the low X-ray luminosity of our systems, their
gas mass fractions are found to be almost universally low
compared to those seen in X-ray selected systems, which typically find $f_{\rm gas} \sim 0.06-0.08$
within our mass range of $10^{13}-10^{14} ~\mathrm{M_{\sun}}$, \citep{SUN09, GONZALEZ13, LIANG16}.
Within our sample, only group 200130 (MKW4) has such a substantial gas reservoir.

\section{Discussion} \label{SDISCUSSION}

\subsection{High Entropy Limits}

The results above indicate that two of our groups have interestingly
high lower limits on the central entropy of their intragroup
gas. However, it is worth asking how realistic these limits are,
especially given the assumptions that have gone into constructing
them. The entropy calculation used here for the detected systems
assumes an isothermal gas with
temperatures determined from the X-ray spectra, and a gas density
profile derived from the deprojected emissivity profile of group
emission. To determine limits for the undetected groups
we instead use the temperature predicted
for each group from their optical luminosity-based masses,
which have some uncertainty from the scatter in the mass-$L_r$
relation used. The uncertainty on mass also affects the shape of the
density profile, since we assume $r_c=0.2r_{500}$, where
$r_{500}$ is also derived from optical luminosity-based
masses. Furthermore, the assumed factor of 0.2, comparable to the mean
ratio of $r_c$ to $r_{500}$ for the rest of the sample, may not be
valid for groups with such high entropy --- high entropy gas will
redistribute itself within the halo, puffing up the intragroup medium
and increasing the core radius. 

It is interesting therefore to see how far we can push these
assumptions before we reach entropy limits comparable to the rest of
the sample. The most obvious question is, how reliably do we
know the assumed temperature? We have already discussed that, for the
detected systems, the 
temperature estimates are, on average, essentially unbiased.
Whilst scatter is large, 0.15 dex, if we decrease the
temperature by 0.15 dex and propagate this change into $r_{c}$,
entropy lower limits decrease to only $>181$ and $>164 ~\mathrm{keV
~cm^2}$ for 100053 and 200099 respectively. Therefore poorly estimated
temperature alone cannot be responsible for the observed high entropy
limits.

The second main assumption involved in our calculation of the entropy
at $10~\mathrm{kpc}$ radius is the form of the gas density profile. The key parameter
here is the value of the core radius, which was assumed to take
the value $0.2 r_{500}$. We therefore examine the fraction of 
$r_{500}$ which would be required to reduce the measured limits at $10~\mathrm{kpc}$ to 
$100 ~\mathrm{keV ~cm^2}$. From our APEC normalisation upper limit,
we find that this core radius would need to drop to $< 0.05r_{500}$ for
both groups, giving entropy profiles which drop sharply at small radius.
This seems highly unlikely for a group with undetectably
low X-ray surface brightness. As we noted earlier, if such a group
contains high entropy gas then the core radius is likely to take
a {\it larger} than normal value.

\subsection{Substructure} \label{SRESSUBSTRCT}

We conclude from the above discussion that the high entropy limits determined
for 100053 and 200099 are unlikely to be reduced substantially by
adjustments to our assumptions about the gas temperature or density.
However, a third possibility remains to avoid the conclusion that
these systems contain high entropy gas. Could it be that the gas
in these systems has not yet been heated, since they are still
collapsing? Our initial dynamical screening was, of course, designed
to eliminate this possibility. Given the improvement in optical data quality 
since the initial selection of our group sample using the G3Cv04 GAMA catalogue,
we now re-visit the question of substructure for
the whole sample using the G3Cv06 data.  
\begin{table}
\centering
	\caption{Substructure in the G3Cv06 Group Sample.} \label{TSUBSTRUCT}
	\begin{threeparttable}
	  \begin{tabular}{lccc}
	  \hline \hline
GroupID	&	$\zeta_\beta$	&	$\zeta_{AST}$	&	$\zeta_{AD}$	\\
\hline
100053	&	1.42 (0.93)	&	1.19 (0.70)	&	2.39	\\
200015	&	1.22 (0.81)	&	0.82 (0.23)	&	1.11	\\
200017	&	1.06 (0.63)	&	1.39 (0.78)	&	3.97	\\
200043	&	1.86 ($>$0.99)	&	2.40 (0.98)	&	1.44	\\
200054	&	0.78 (0.26)	&	1.24 (0.69)	&	3.41	\\
200099	&	3.08 ($>$0.99)	&	2.08 (0.95)	&	2.19	\\
200115	&	1.43 (0.89)	&	1.42 (0.76)	&	1.03	\\
200130	&	1.78 (0.99)	&	1.23 (0.74)	&	3250	\\ 
300008	&	1.11 (0.68)	&	0.79 (0.21)	&	1.03	\\
300067	&	1.55 (0.94)	&	1.93 (0.93)	&	2.86	\\
\hline
    \end{tabular}
     \begin{tablenotes}
      \item $\zeta_\beta$, $\zeta_{AST}$ and $\zeta_{AD}$ show the mirror symmetry, angular separation test and velocity non-normality substructure indicators as described in Section \ref{SDATA}. The numbers in parentheses show the significance of any
observed substructure for the first two tests.
     \end{tablenotes}
    \end{threeparttable}
\end{table}

In Table \ref{TSUBSTRUCT} we show the recalculated substructure
statistics of our group sample using the deeper data now available and
centred on X-ray emission where possible.  We remind the reader that
the original G3Cv04 substructure thresholds were $\zeta_{\beta} <
1.9$, $\zeta_{AST} < 1.68$ and $\zeta_{AD} < 1.82$. 
Our target groups all fell below these threshold values using the
G3Cv04 galaxy data. Due to the updated
group catalogue a rigorous comparison of the new substructure statistics
to the original, carefully calibrated, thresholds is difficult, and
would require recalibration of these thresholds using mock datasets
constructed to match the deeper data. Qualitatively though, such a comparison 
has a number of interesting implications.

\begin{figure}
\centering
	\includegraphics[width=0.49\textwidth]{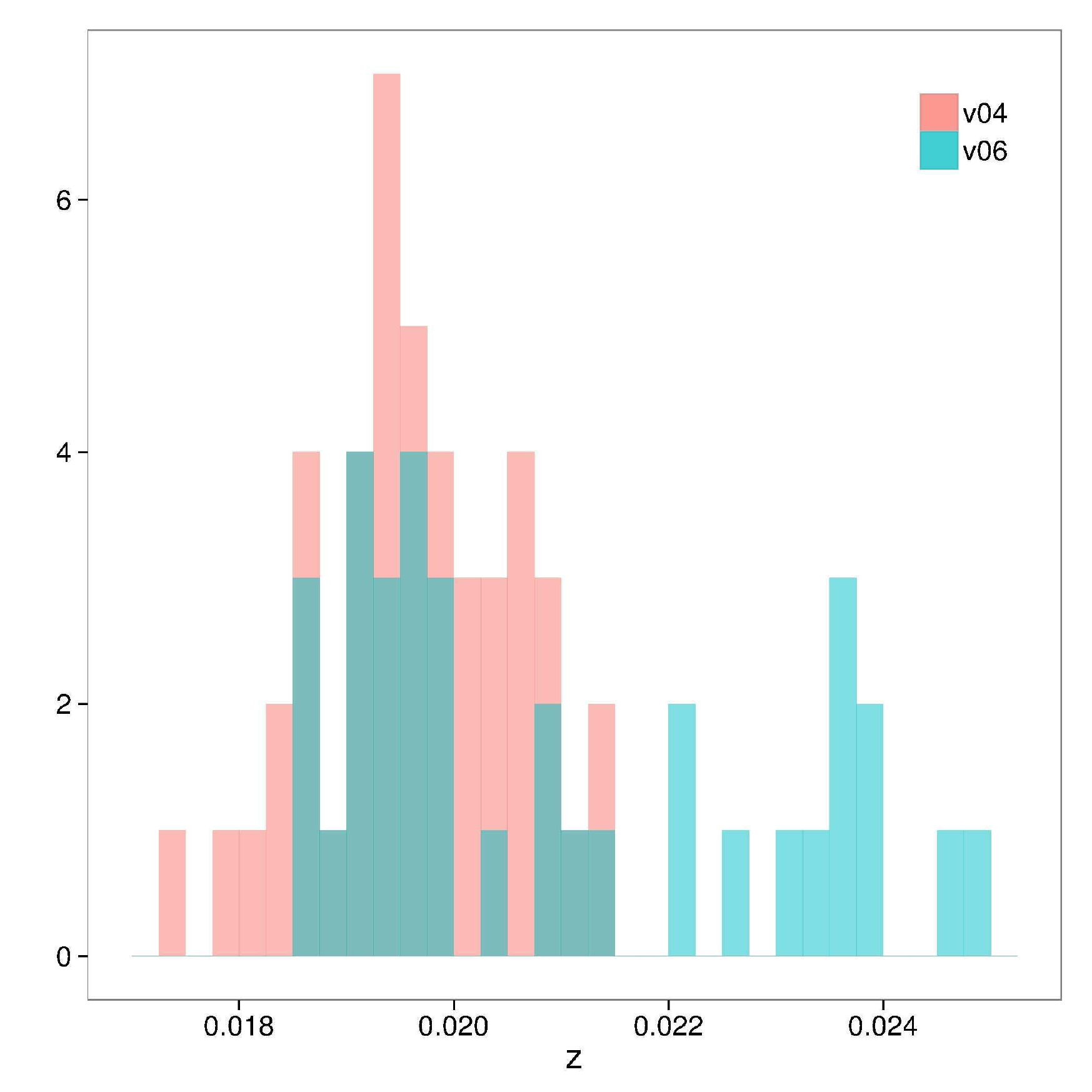}
	\caption{Histogram of galaxy redshifts for members of group 200130 (MKW4) identified within G3Cv04 (red) and G3Cv06 (blue) group catalogues. A large fraction of the galaxies linked to this group in G3Cv04 are part of another group in G3Cv06 (see text and Section \ref{SGRPSAMPLE}).}
	\label{FMKW4Z}
\end{figure}

The most apparent change is that the Anderson-Darling test has become
much more discriminating.  Obviously, the inclusion of more galaxies
will result in any deviation from normality in the velocity
distribution becoming more significant, so this should not surprise
us. However, the value of $\zeta_{AD}=3250$ for 200130 (MKW4)
indicates that the revised selection has introduced a major
perturbation in its velocities. Further examination indicates that
a second structure, centred at $z \sim 0.0235$, has been linked
into the group in the G3Cv06 catalogue, causing the group velocity histogram to be
significantly skewed, as shown in Figure \ref{FMKW4Z}. At the same time, some
of the galaxies in the G3Cv04 group have been lost, as discussed at the end
of Section~\ref{SGRPSAMPLE}.

In contrast, the two spatial substructure tests do not show major
changes from G3Cv04 to G3Cv06 groups, with 
only a few showing more substructure than the
original limits. The mirror symmetry gives a significant ($\ge
95$ per cent) substructure result for groups 200043, 200099 and 200130. The
angular symmetry test similarly highlights groups 200043 and 200099.
The significant mirror symmetry result for 200130 is not
surprising given its intersection with the survey edge.

Comparing the substructure indices of the two non-detected groups,
with the others, we see they have similar, moderate values of $\zeta_{AD}$ 
(ranking 5th and 6th highest in the sample). For the spatial tests, the
values for group 100053 are low, and 
typical of the rest of our sample, suggesting that it is a
virialised system. In contrast, group 200099 shows some of the highest
substructure in the sample (ranked 1st for  $\zeta_{\beta}$ and second
for $\zeta_{AST}$). However, this may be related to the uncertainties in
centring already noted, since 200099 has no obvious bright central
galaxy. Recalculating substructure around a simple centre-of-light
centroid, without iterating, we observe considerably less projected
substructure, with $\zeta_\beta=1.47 ~(\mathrm{p}=0.94)$ and $\zeta_{AST}=1.42
~(\mathrm{p}=0.81)$. Whilst the lack of a dominant central galaxy could
result from incomplete virialisation of the group, we note
that if it is a high entropy group then gas will be inhibited from cooling
onto any central galaxy and fuelling growth through late star formation.

On the basis of substructure, we therefore conclude that we have 
no strong evidence that the
groups 100053 and 200099 are not real, collapsed structures. This
suggests that the observed high entropy limits provide genuine
indicators of the entropy of the gas in these systems.

\subsection{Mass Estimation Quality} \label{SRESMASSES}

The ability to estimate masses is important for this work.
It was the basis for the predicted X-ray fluxes which were used
in selecting the sample for {\it Chandra} observation and is also
used in the calculation of the group entropies. 
We have used the X-ray temperature to derive our mass estimates
wherever possible, as described in Section~\ref{SXSPEC},
and as discussed in Section~\ref{SRESLT} above, we believe our predicted
temperatures, and therefore mass estimates, are in general unbiased,
though highly scattered. 

Nonetheless, it is of considerable interest to compare the results
from different mass estimation techniques applied to our sample.
A variety of mass estimates can be derived from analysis of the
group galaxies using their dynamics, optical luminosity or spatial
distribution. For groups which are simple, relaxed and virialised, 
and which have typical star formation efficiency, we would expect
to obtain good agreement between these different optical mass estimates.
So, major differences between results from the different estimators
may indicate groups which are unrelaxed or atypical. It is therefore
highly relevant to examine whether this is the case for our two
X-ray undetected groups.

The optical mass estimators we employ are mostly taken from the
study of \citet{PEARSON15}, and we therefore
reselect the member galaxies for each of our
groups using the method employed by these authors. Galaxies are
extracted from a cylindrical volume with projected radius of 1 Mpc and
velocity depth $\pm 3 \sigma$ along the line-of-sight centred on the X-ray centroid where possible, where velocity
dispersion $\sigma$ is derived using the gapper estimator of
\citet{BEERS90}. We refer to this galaxy sample as the volumetric
group sample, in contrast to the GAMA Friend-of-Friends sample defined
by \citet{ROBOTHAM11}. Using the volumetric sample, we construct a
series of mass estimates based on (i) the observed group richness and
(ii) luminosity (both extrapolated from the $m_r = 19.8 ~\mathrm{mag}$ flux limit to
a standard limiting absolute magnitude $M_r = -16.5 ~\mathrm{Mag}$ assuming 
a cluster luminosity function derived from the SDSS \citep{POPESSO05}),
(iii) galaxy and (iv) luminosity overdensity (from fits to a 
NFW radial density profile \citep{NFW96}), 
and (v) a dynamical mass estimator
$\propto \sigma^{3\alpha}$. These mass estimates, labelled $M_N$,
$M_L$, $M_\delta$, $M_{\delta_L}$ and $M_{\sigma}$ respectively, are
based on mass-proxy relations that have been calibrated against
$M_{500}$ from a sample of X-ray selected groups, as described in
\citet{PEARSON15}. 

In addition to these five galaxy-based mass 
estimators, we also include masses derived from GAMA FoF
total light ($M_{L,GAMA}$, Section \ref{SDATA}) and GAMA group masses
derived from the FoF group velocity dispersion \citep[$M_{\sigma,
GAMA}$, ][]{ROBOTHAM11}, calibrated on the GAMA mock data and
applied to the G3Cv06 GAMA catalogue. 
\begin{table*} \begin{minipage}{\textwidth} \centering
	\caption{$M_{500}$ estimates (in units of $10^{13} \mathrm{M_{\sun}}$, assuming $h_{70}=1$) for our group sample. }
	\label{TMASSES}
	\begin{threeparttable}
	  \begin{tabular}{lcccccccc}
	  \hline \hline
GroupID	&	$M_{L, GAMA}$	&	$M_{\sigma, GAMA}$	&	$M_N$	&	$M_L$	&	$M_{\delta}$	&	$M_{\delta_L}$	&	$M_{\sigma}$	&	$M_{\rm X\mhyphen ray}$\tnote{a}	\\
\hline
100053	&	5.91	&	6.46	&	3.84	&	6.40	&	2.50	&	2.46	&	1.06	&	(-)		\\
200015	&	4.79	&	6.61	&	14.06	&	9.78	&	24.8	&	28.1	&	6.49	&	($3.16\pm0.83$)		\\
200017	&	4.01	&	7.71	&	2.90	&	3.95	&	6.44	&	5.83	&	5.07	&	($4.74\pm2.33$)		\\
200043	&	5.25	&	8.06	&	5.97	&	11.9	&	7.67	&	10.9	&	5.24	&	($2.77\pm1.47$)		\\
200054	&	4.61	&	8.48	&	6.91	&	9.34	&	6.27	&	7.83	&	6.00	&	($16.2\pm7.53$)		\\
200099	&	4.73	&	3.32	&	1.83	&	3.32	&	0.90	&	0.91	&	1.44	&	(-)		\\
200115	&	2.62	&	4.50	&	0.45	&	0.54	&	2.40	&	1.55	&	1.63	&	($1.20\pm0.12$)		\\
200130	&	7.22	&	13.1	&	4.17	&	7.15	&	8.05	&	3.91	&	9.89	&	($7.66\pm0.15$)		\\
300008	&	3.60	&	8.88	&	2.89	&	3.98	&	6.43	&	4.22	&	4.22	&	($6.80\pm1.83$)		\\
300067	&	2.63	&	3.03	&	1.21	&	1.00	&	1.27	&	0.95	&	1.68	&	($2.44\pm0.44$)		\\

\hline
    \end{tabular}
     \begin{tablenotes}
	\item[a] For comparison, X-ray $M_{500}$ as reported in Table \ref{TRESULTS}.
     \end{tablenotes}
    \end{threeparttable}
\end{minipage}
\end{table*}

\begin{figure}
\centering
	\includegraphics[width=0.49\textwidth]{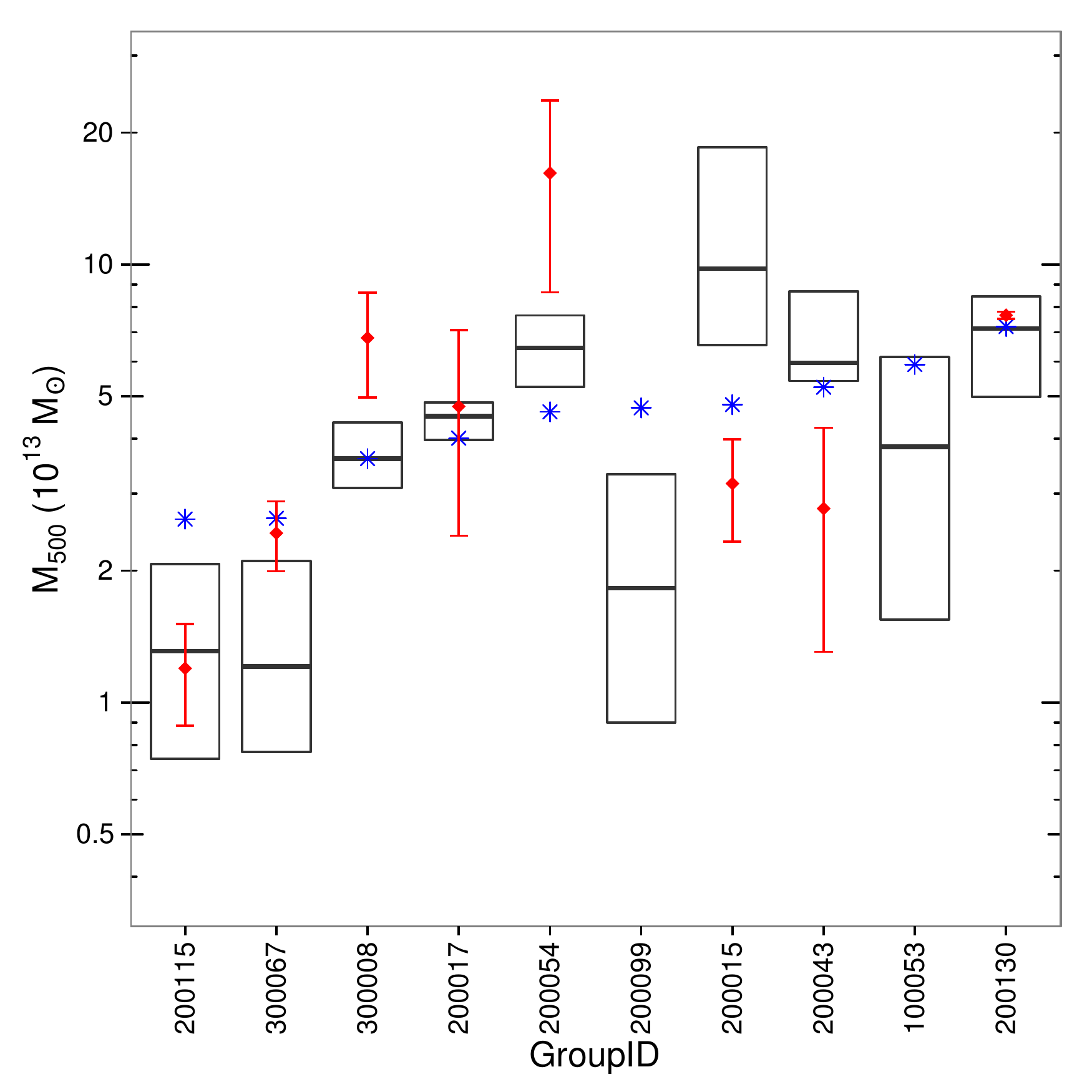}
	\caption{Distribution of predicted group $M_{500}$ ordered by the GAMA luminosity predicted
	mass. Included in the boxplot are all mass estimates from Table \ref{TMASSES} with the 
	exception of X-ray based mass estimates.  The \textit{blue stars} are the predicted masses 
	used for the X-ray feasibility (derived from the GAMA luminosity), the \textit{red diamonds} 
	are the $M-T$ based masses of Table \ref{TRESULTS}. The boxes represent the median and 25th 
	to 75th quartiles of mass estimates.}
	\label{FMBOX}
\end{figure}

In Table \ref{TMASSES} and Figure \ref{FMBOX}  
we summarise these mass estimates, and compare them with the
X-ray derived masses for the eight detected systems. 
The boxplot in Figure \ref{FMBOX} shows the
spread of mass estimates according to their interquartile range for
each group. With only 7 masses going into each boxplot, the statistics here are limited,
so definitive conclusions are difficult to draw. Nonetheless, a number of
interesting points can be noted.

As seen in Figure \ref{FMBOX}, of the eight groups with detected X-ray
emission we see four groups (200115, 300067, 200017 and 200130)
for which the X-ray derived masses are
consistent (within $1\sigma$) with the interquartile range (IQR) of
mass estimates. In the other four cases, the X-ray mass lies within about
a factor 2 of the interquartile box -- above it in two cases, and below 
it in two others.

The masses ($M_{L,GAMA}$) predicted from the GAMA group optical
luminosity were used to estimate the expected X-ray properties
when selecting our sample for observation with {\it Chandra}.
Comparing these masses (blue stars in Figure~\ref{FMBOX}) with
the X-ray masses for the eight detected systems,   
we note that only one group (200115) had its mass overpredicted
by more than a factor of two. Since our substructure screening was estimated (Section~\ref{SSUBCAL})
to leave a 6 per cent chance for a given group to have its mass overestimated
by a factor of 2 or more, a binomial calculation shows that there
is a 34 per cent chance for a sample of 10 groups to contain
one group whose predicted properties are significantly overestimated. 
So, we should not be too surprised to find one case in the sample.
Note, however, that
the X-ray derived mass for this system is contained well within the
IQR of the whole set of optical mass estimators for this group.

Comparing the size of the IQR for each group, we see that the two
groups with no diffuse X-ray emission also have the broadest IQR of
the sample. Examining individual mass estimates, we see that both
the dynamical, $M_{\sigma}$, mass for these groups and the overdensity
based masses, $M_\delta$ and $M_{\delta_L}$, are low compared to
masses estimated from a simple count of light or number. 
Since the overdensity masses are derived by fitting the radial profiles
of the group galaxies, the low values of these estimates suggest that
these groups may be characterised by small overdense cores
located within larger collapsing structures. This interpretation
is strengthened by the fact that a similar discrepancy is observed
between the overdensity and richness mass estimates for the group
300067. As already noted, this group shows little diffuse emission other
than a concentrated halo of emission around the central bright galaxy.

\section{Conclusions} \label{SCONCLUSION}

This study has investigated the hot gas properties of a small sample
of galaxy groups selected to have good galaxy membership data from the
GAMA survey and to show little optical substructure. Using data from 
the \textit{Chandra} Observatory we detect hot intergalactic gas
in eight of our ten groups, and estimate X-ray temperatures,
luminosities and central gas entropies, searching for evidence of the
existence of groups with the high gas entropies predicted by some
preheating models.  Two of our groups are high entropy candidates,
with $2\sigma$ lower limits on central gas entropy which lie
at the upper edge of the $1\sigma$ range predicted by the OWLS AGN feedback model.  All other
groups have entropies lying within the range found in X-ray selected group samples, but with a median value shifted towards higher entropy, 
consistent with the reduced X-ray luminosity and gas fractions observed within our sample.

We have examined the robustness of our entropy limits against
uncertainties in temperature or core radius, and conclude
that these are unlikely to reduce our high upper limits into
the range seen in normal X-ray bright systems.

Two different approaches have been applied to investigate the possibility
that these two high entropy candidates may, notwithstanding our substructure
screening, be uncollapsed or unvirialised systems. A closer examination
of the substructure statistics, using the deeper GAMA data which became available
since our initial sample was selected, does not reveal any convincing evidence
for the two undetected groups being uncollapsed or unvirialised. 
However, our second test, comparing the mass estimates for the sample
derived from a basket of galaxy-based mass estimators, does produce 
evidence that the high entropy candidates may be `special'. These
are the two groups from our sample which show the largest
discrepancies between the different galaxy-based mass estimates.
In both cases, masses derived from the galaxy overdensity profiles
or the velocity dispersion are substantially lower than those
derived from richness or total group $r$-band luminosity.
One possible interpretation which we suggest, is that these
two groups are not strongly substructured or asymmetrical
(hence the low values from the substructure statistics) but 
are not yet fully collapsed.

Whether this possibility could account for the lack of 
detectable diffuse X-ray emission in these groups without the need
to invoke high entropy gas is not clear. In the absence of
strong preheating, a collapsed core can still
generate significant X-ray emission, as is seen in the case of
group 300067. A definitive answer to the question of whether
groups 100053 and 200099 contain high entropy gas requires 
deeper X-ray observations able to detect the hot gas. 
It should be noted that these two systems are among
the three shortest {\it Chandra} exposure times in our study.

In addition to the possibility of high entropy gas within two of our
groups, it is clear that our optically-selected sample deviates
significantly from the properties of typical X-ray selected groups. 
The raised gas entropies and lower X-ray luminosities and gas fractions seen in our sample
highlight the importance of selection effects when studying the 
properties of gas within collapsed groups.

Future deep X-ray studies of a larger sample of groups selected
from the more recent GAMA data releases should provide strong
constraints on the true distribution gas properties in collapsed
groups independent of X-ray selection effects. This distribution can
then be used to discriminate between competing models for the cosmic
feedback arising from AGN and supernovae.  In this context, the
present study should be viewed as a pathfinder.

\section{Acknowledgements}

RJP acknowledges the support of an STFC Postgraduate Studentship. PN and RGB acknowledge the support
of the Science and Technology Facilities Council (ST/L00075X/1 and ST/P000541/1). PN also acknowledges
the support of the Royal Society through the award of a University Research Fellowship. AB acknowledges support from NSERC (Canada) 
through the Discovery Grant program and to the Pauli Center for Theoretical Studies ETH UZH.
He would also like to thank University of Zurich's Institute for Computational Sciences,
and especially the members of the Institute's Center for Theoretical Astrophysics and Cosmology,
for their hospitality during his recent extended visit.  IGM is supported by a STFC
Advanced Fellowship. We would also like to thank
Alastair Sanderson and Ewan O'Sullivan for their assistance when
selecting our groups, and an anonymous referee for valuable comments on the paper.

GAMA is a joint European-Australasian project based around a spectroscopic campaign using the
Anglo-Australian Telescope. The GAMA input catalogue is based on data taken from the Sloan
Digital Sky Survey and the UKIRT Infrared Deep Sky Survey. Complementary imaging of the GAMA 
regions is being obtained by a number of independent survey programmes including GALEX MIS, 
VST KiDS, VISTA VIKING, WISE, Herschel-ATLAS, GMRT and ASKAP providing UV to radio coverage.
GAMA is funded by the STFC (UK), the ARC (Australia), the AAO, and the participating institutions.
The GAMA website is \url{http://www.gama-survey.org/}.


\bibliographystyle{mn2e}
\bibliography{bib}

\end{document}